\documentstyle[preprint,aps]{revtex}
                    \tighten
                    \begin{document}
                    \title{Quantum Ballistic Evolution in Quantum
          Mechanics: Application to Quantum Computers}
                    \author{Paul Benioff\\
                     Physics Division, Argonne National Laboratory \\
                     Argonne, IL 60439 \\
                     e-mail: pbenioff@anl.gov}
                    \date{April 24, 1996}

                    \maketitle
                    \begin{abstract}  
          Quantum computers are important examples of processes whose
          evolution can be described in terms of iterations of single step
          operators or their adjoints. Based on this, Hamiltonian evolution
          of processes with associated step operators $T$ is investigated
          here.  The main limitation of this paper is to processes which
          evolve quantum ballistically, i.e. motion restricted to a
          collection of nonintersecting or distinct paths on an arbitrary
          basis. The main goal of this paper is proof of a theorem which
          gives necessary and sufficient conditions that $T$ must satisfy
          so that there exists a Hamiltonian description of quantum
          ballistic evolution for the process, namely, that $T$ is a
          partial isometry and is orthogonality preserving and stable on
          some basis.  Simple examples of quantum ballistic evolution for
          quantum Turing machines with one and with more than one type of
          elementary step are discussed.  It is seen that for
          nondeterministic machines the basis set can be quite complex with
          much entanglement present.  
          It is also proved that, given a step operator $T$ for an
          arbitrary {\em deterministic} quantum Turing machine, it is
          decidable if $T$ is stable and orthogonality preserving, and if
          quantum ballistic evolution is possible.  The proof fails if $T$
          is a step operator for a {\em nondeterministic} machine.  It is
          an open question if such a decision procedure exists for
          nondeterministic machines. This problem does not occur in classical 
          mechanics.  Also the definition of quantum Turing machines used here
          is compared with that used by other authors.
          \end{abstract}

          \section{Introduction}
          There are many processes in physics which can be described in
          terms of a sequence of steps. The computation process furnishes
          many examples. Each computer program is a collection of
          elementary steps which the physical system (computer) undergoes. 
          Computation using a program on a given input consists of
          iteration of the program steps where the particular elementary
          steps carried out at the n+1st iteration depend on the system
          state and the available elementary steps.  Each elementary step
          is local in the sense that changes in a space region depend on
          conditions in the region and not on far distant conditions.

          Each computation by a given program on a specified input yields a
          sequence or path of distinct computation states.  A computation
          state is a complete global specification of the states of all
          relevant degrees of freedom of the system carrying out the
          computation.  For a Turing machine a computation state specifies
          the string of bits or numbers on the computation tape and the
          internal state and position of the scanning head.  For networks
          of gates a computation state is a complete specification of the
          states of the systems moving in the wires and the states of the
          gates in the network.  

          The collection of all paths generated by the given program acting
          on all possible inputs is a collection of finite and infinite
          paths of computation states. The collection of paths depends on
          the computer program being considered as it is different for
          different programs.  In many computational models, paths
          representing halting or nonhalting computations are finite or
          infinite respectively. In other models all paths are infinite
          with halting computations specified by a system flag.

          As is well known \cite{Landauer,Bennett} computations can be
          reversible or irreversible.  Irreversible computations are those
          for which each computation state has at most one successor but
          may have more than one predecessor.  Computation states in
          reversible computations have at most one successor and one
          predecessor.

          Since the interest here is in Hamiltonian models of process such
          as computation, consideration is limited to reversible
          computations only.  As Bennett \cite{Bennett} has shown, this is
          not a limitation in that for each Turing machine computation
          (reversible or irreversible) there is an equivalent reversible
          computation which is slower and has more relevant degrees of
          freedom (history and copying degrees) than the original
          computation.

          This work, which showed that computation could be performed by
          reversible or information preserving steps only, along with that
          of Landauer \cite{Landauer} formed the basis for early work
          \cite{Benioff} on quantum mechanical Hamiltonian models of
          computers as Turing machines.  This work, along with that of
          Feynman \cite{Feynman} and Deutsch \cite{Deutsch}, formed the
          basis for the recent blossoming of the field.  Recent work
          includes that of Lloyd on the halting problem for quantum
          computers  \cite{Lloyd1}, and Lloyd \cite{Lloyd2} and DiVincenzo
          \cite{Divincenzo} and others \cite{DBE,Bar} on the universality
          of 2-bit quantum gates for quantum circuit computation. The work
          of Shor \cite{Shor}, showing that the integer factoring problem
          could be solved in polynomial time on a quantum computer, has
          provided much of the impetus for the recent work.

          Work has also been proceeding on developing physical models of
          quantum computers.  As is well known proper functioning of a
          quantum computer requires that phase relations be maintained
          between the component states of all the degrees of freedom in the
          model. Landauer \cite{Landauer1} has repeatedly emphasized the
          problem of physical realization of quantum computation in that
          environmental noise and decoherence cause degradation of
          performance.  Additional work on the effects of decoherence on
          quantum computation has been done by Unruh \cite{Unruh} and
          others \cite{C-L-S-Z}.  Recent work on quantum error correcting
          codes \cite{Cald-Shor,Shor1,PGCZ} gives hope that the effects of
          noise and decoherence can be minimized.  It is also clear that it
          is advantageous to minimize the number of degrees of freedom
          needed to carry out a quantum computation since fewer degrees
          involved means less effort is needed to minimize environmental
          influences.

          The work of this paper is based on a translation of step
          processes such as reversible computation into quantum mechanics. 
          To each process is associated a bounded linear operator $T$,
          called a "step operator" for the process such that iteration of
          powers of $T$ (or its adjoint $T^{\dag}$ model successive steps
          forward (or backward) of the process.  Note that no association
          of a finite time interval with a step is assumed.  As a result
          step operators can be used directly in the construction of
          Hamiltonians.  Additional discussion is provided in the sext
          section.

          The states of the model process system can be represented as
          states in some basis.  For the purposes of this paper it is not
          necessary that $T$ be equal to a sum of elementary step
          operators.  However, if the process consists of a set of
          elementary steps, such as a model of a computer program, then it
          is useful to write $T$ as a sum of corresponding elementary step
          operators.  

          The procedure used here differs from that used by  Deutsch
          \cite{Deutsch} and Bernstein and Vazirani \cite{BV} in their
          description of quantum computation.  They assumed that the step
          operator is unitary, spatially local, and associated with a
          finite time interval.  In this case a Hamiltonian can be defined
          by $T=e^{-iHt}$.  

          Because of problems with this approach, such as
          the nonexistence of a Hamiltonian which is simple and local, this
          approach is not used here.   Instead a step operator $T$ for a 
process such as quantum computation is associated with an infinitesimal
 time interval. As a result, it can 
be used to construct a Hamiltonian $H(T)$ which describes the unitary time 
evolution of the process.  The Hamiltonian is time independent, selfadjoint, 
and has the complexity of $T$ and not of all steps of the process.
The step operator
          $T$  associated with a process need not be unitary,
          selfadjoint, or even normal.   More details on a comparison between 
this approach and that of Deutsch and Bernstein and Vazirani \cite{Deutsch,BV}
will be discussed later on.

In earlier work \cite{Benioff1,Benioff2} unitary step operators for quantum 
Turing machines were constructed.  However the work was limited to 
deterministic computations only.  In addition the unitarity was artificial in 
that it held only 
for the subspace of states defined by all iterations of the step operator and 
its adjoint.  In general (e.g. for universal machines) this subspace cannot be 
defined effectively.  Also on the larger space of all states (which can be 
effectively defined) the step operator 
was not unitary and its properties on the larger space were not considered.

          In this paper much attention is given to conditions that a step
          operator $T$ must satisfy such that iteration of $T$ or its
          adjoint on the states in some basis generates a collection of
          paths in the basis.  A path is a finite or infinite sequence of
          distinct states in a basis. A basis is a set of pairwise
          orthogonal, normalized states which span a Hilbert space.  The
          idea is that if the model process system is in any state in a
          path, then successive states in the path represent successive
          steps of the process.  
 
       An additional requirement is that
        the model process step operator be distinct path
          generating in some basis.  This follows the restriction made for 
        reversible classical computations \cite{Bennett}. This means that if 
the process is started on different input states, the paths generated by 
successive steps of the process must be distinct and have no overlap.  
Otherwise computations started on distinct inputs would overlap and one would 
not know which input was associated with the output.
         
          In general an arbitrary operator as a candidate model step
          operator may not have any of these desired properties.  Iteration
          of the operator may not generate a path in any basis in that
          orthogonality of states is not preserved under iteration.   Or
          the operator may generate paths which branch and join into a
          network of interconnected paths in all bases.  Note that in order to ensure
          that no paths join or branch it is necessary  that the states in
          all the paths be in the same basis.  For example if each of two
          different paths were in different bases, then the overlap of
          states in the different bases would destroy the distinctness of
          the paths. If there were no overlap between the states in the two
          paths, then another basis, which includes the states in both of
          the paths, would be a suitable basis for the requirement of
          distinct path generation.

          The main goal of this paper is to give necessary and sufficient
          conditions that  an arbitrary model process step operator must
          have in order that there exist a quantum mechanical Hamiltonian
          which describes quantum ballistic evolution of the process. 
          Quantum ballistic evolution refers to the "motion" of the model
          process system along paths of states and is limited to step
          operators which are distinct path generating.   The paths are
          defined by iteration of the process step operator  or its adjoint
          on states in the basis.    Under this type of evolution any wave
          packet of states on a path moves along the path, spreading out as
          it moves.
           
          It is important to note that full advantage is taken here of the
          fact that in quantum mechanics there exist many inequivalent
          basis sets.  In classical mechanics there is only
          one\footnote{This distinction between classical and quantum
          mechanics follows from the fact that bases are in 1-1
          correspondence with maximally fine resolutions of the identity
          (i.e. those in which all the projection operators are one
          dimensional).  In quantum mechanics there are many maximally fine
          resolutions which do not commute.  In classical mechanics there
          is only one. Here inequivalent bases are defined to be those for
          which the corresponding resolutions of the identity do not
          commute.}.  In particular nondeterministic computations, which
          allow arbitrary bit transformations (such as $\vert i\rangle 
\rightarrow \alpha \vert 0\rangle +\beta \vert 1\rangle$ for $i=0,1$), and 
deterministic computations, which limit bit transformations to $0\rightarrow 
1, 1\rightarrow 0$ only, are included.  This is done by allowing basis
          sets, with respect to which the paths are defined and quantum
          ballistic evolution occurs, to be arbitrarily complex with
          entanglements between the component model systems.

          Another goal of this paper is to determine if there exists an
          effective decision procedure to decide if a model step operator
          for an arbitrary process is distinct path generating and  thus if
          a Hamiltonian description of quantum ballistic evolution exists.
          It will be seen that in general such a decision procedure does
          not exist.  However, for models of computation, this problem
          exists only for nondeterministic computations.  An effective
          decision procedure is shown to exist for deterministic models. 

          In the next section there is more discussion on paths, distinct
          path generation by process step operators, and quantum ballistic
          evolution for partial isometries. Feynman's prescription
          \cite{Feynman} of construction of a Hamiltonian from process step
          operators is introduced.  

          In Section \ref{ipo} definitions of stability and orthogonality
          preserving for operators are introduced.  Some theorems are
          stated and proved including the result that stability plus
          orthogonality preservation are equivalent to distinct path
          generation.  Section \ref{ppi} introduces power partial
          isometries and gives some of their main properties of interest
          here.  The equivalence between complete orthogonality
          preservation and power partial isometry is proved.

          Section \ref{main} contains the main result. It is proved that
          necessary and sufficient conditions for the existence of a
          Hamiltonian description of quantum ballistic evolution on some
          basis for a process step operator $T$ is that $T$ is a partial
          isometry which is orthogonality preserving and stable.  Canonical
          eigenfunctions and eigenvalues are given for the Feynman
          Hamiltonian (Eq. \ref{ham}). 

          Sections \ref{examples} and \ref{qctm} give illustrative examples
          and a definition of the step operators for Turing machine models
          of quantum computation.  The existence of nondeterministic
          quantum Turing machines which evolve quantum ballistically is
          shown by explicit example construction.  Following is
          a discussion of effective determination of the existence of a
          Hamiltonian description of quantum ballistic evolution for a
          process with associated step operator $T$. Some other aspects,
          including a discussion of the approach used here and that of
          Deutsch and Bernstein and Vazirani \cite{Deutsch,BV}, are
          included in Section \ref{discussion}.


          \section{Quantum Ballistic Evolution}
          \label{step-qbe}
          As noted the interest here is in step operators $T$ which are
          distinct path generating.  That is  \\
          \\
          {\bf Definition 1} {\em A step operator $T$ and its adjoint are
          {\em distinct path generating on a basis $B$} if iterations of
          $T$ and $T^{\dag}$, started at any state in $B$, generate paths
          in $B$ that are distinct in that they do not intersect or join
          one another.}\\
          \\ 
          In more precise terms this definition means that $T$ and its
          adjoint are distinct path generating if for all states $\vert
          p_{j}\rangle$ in a basis $B$, if $T\vert p_{j}\rangle \neq 0$,
          then there exists a unique state $\vert p_{k}\rangle$ in $B$ such
          that $T\vert p_{j}\rangle =\vert p_{k}$ and $T^{\dag}\vert
          p_{k}\rangle =\vert p_{j}\rangle$.  A similar statement holds
          with $T^{\dag}$ replacing $T$.  Note that distinct path generation is 
defined here for norm preserving motion only.  In future work this restriction 
will be removed from the definition by allowing $T\vert p_{j}\rangle 
=\alpha_{jk} \vert p_{k}\rangle$ where $\alpha_{jk}$ can be different from $1$. 

          The purpose of this definition is to get rid of paths which join,
          branch or intersect.  Note that paths of $0$ length are included.
          (i.e. for some $\vert p_{j}\rangle, T\vert p_{j}\rangle =
          T^{\dag}\vert p_{j}\rangle =0$).  Also it is clear that any $T$
          satisfying the definition is a partial isometry.  An operator is
          a partial isometry if both $T^{\dag}T$ and $TT^{\dag}$ are
          projection operators.

          Step operators $T$ which are distinct path generating on some
          basis can be used to model the evolution of some system whose
          elementary steps are modelled by $T$ and $T^{\dag}$.  The adjoint
          is used instead of the inverse as $T$ may not have an inverse.  A
          state of the system on a path consists of a wave packet of states
          on a path.  In general such a wave packet has the form 
          \begin{equation}
          \Psi(t) = \sum_{n=0}^{M} c_{n}(t)T^{n}\vert 0\rangle
          +\sum_{n=1}^{L}c_{-n}(t)(T^{\dag})^{n}\vert 0\rangle.
          \label{packet}
          \end{equation}
          The time development is shown in Eq. \ref{packet} by the explicit
          time dependence of the coefficients $c_{n}(t)$. 

          Quantum ballistic evolution refers to the time evolution of such
          packets along distinct paths in some basis.  As these packets
          move under the action of some Hamiltonian they spread out along
          the paths.  If a path is two way infinite, $M=L=\infty$, motion 
continues with no momentum change.  If a path is infinite in one direction only
          reflection occurs at the path end.  If the path is finite with
          distinct ends, reflection occurs at both ends.  For cyclic paths
          the packet moves around the cycle with interference occurring as
          the packet spreads over a distance greater than one cycle.
          Additional details on the packet spreading are given in
          \cite{Benioff1,Benioff2}. 

          A very simple example of quantum ballistic evolution would be
          free system motion in a collection of quantum wires, which do not
          intersect, branch or join, on a three dimensional space lattice.
          This includes straight or curved lines or line segments, closed
          loops such as circles, chains of closed loops, etc.. For this
          example the basis $B$ is the set $\{\vert x,y,z\rangle \}$ of
          position vectors on the lattice.  In this paper the basis set is
          not fixed in advance, and the character of $B$ on which quantum
          ballistic evolution occurs, if it is possible, is determined by
          the properties of the operator $T$.  

          A schematic representation of wave packets on infinite paths is
          shown in Figure 1.  Here the states in some basis are represented
          as points in space. Two paths, shown by dotted lines are shown. 
          Path A is nonterminating and path B has a terminus T.  Two wave
          packets, $\psi_{1}(t)$ and $\psi_{2}(t)$ are shown. For each the
          time dependent coefficients are
          represented schematically by $c(t)=r(t)e^{i\theta(t)}$.  The
          basis state dependence of the coefficients is also indicated.  

          Quantum ballistic evolution can also be used to describe motion
          for an initial state which is a linear superposition of wave
          packets in many different paths.  By linearity the packets in
          each path evolve independently of the others.  However, this can
          lead to great entanglement among the different degrees of freedom
          in the system being modelled.  How much entanglement, if any,
          depends on the basis set used and the system being modelled.  In
          quantum computation use of linear superpositions in this manner
          is referred to as computation by quantum parallelism
          \cite{Deutsch}.  For example, a linear combination of the two
          packet states shown in Figure 1 would evolve in parallel.

          A Hamiltonian description of quantum ballistic evolution can be
          obtained by use of Feynman's prescription \cite{Feynman}.  That
          is, given an arbitrary step operator $T$ for a process (which may
          or may not be distinct path generating) define the corresponding
          $H$ by 
          \begin{equation}
          H=K(2-T-T^{\dag}) \label{ham}
          \end{equation}
          This Hamiltonian has the advantage that is it simple.  That is,
          it has a complexity of the order of $T$.  In particular, it does
          not have the complexity of all paths generated by iteration of
          $T$ or its adjoint. In the case that $T$ models a quantum
          computation, $H$ has the complexity of the computer program. This
          is especially desirable in the case that $T$ models a universal
          Turing machine. 

          In general for an arbitrary step operator, the Hamiltonian given
          above does not describe quantum ballistic evolution. If $T$ is a
          step operator for a collection of paths that intersect or join,
          then $e^{-iHt}$ will describe the unitary evolution of some
          process. But it may be a different process from that obtained by
          iteration of $T$.  This is especially the case if iteration of
          $T$ describes an irreversible process.  For example, there are
          operators $T=T_{1}+T_{2}$ such that the Hamiltonian of Eq.
          \ref{ham} describes the evolution of a different process, namely,
          that with the step operator $X=T_{1}+T^{\dag}_{2}$.

          If $T$ is distinct path generating, then the Hamiltonian of Eq.
          \ref{ham} describes quantum ballistic evolution.  To see this,
          consider the power series expansion of $e^{iKt(T+T^{\dag})}$
          where $e^{-iHt} =e^{-2iKt}e^{iKt(T+T^{\dag})}$.  Each term in the
          expansion has the form  $\cdots (T^{\dag})^{m_{4}}T^{m_{3}}
          (T^{\dag})^{m_{2}}T^{m_{1}}$ where the $m_{i}$ are nonnegative
          integers. Because $T$ is distinct path generating, the term will
          describe motion back and forth along each path.  For finite paths
          of length $n$, terms for which $m_{i}\leq n$ for some $i$, will
          give $0$ operating on any state in the path.  If $\psi$ is a
          state not on any path of $T$ then $e^{-iHt}\psi=e^{-2Kit}\psi$
          which shows no change occurs.  

          This is clearly a description of quantum ballistic evolution as
          motion is restricted to be in either direction along the distinct
          paths.  No motion occurs on states outside the paths.  Conversely
          if $H$ is an arbitrary Hamiltonian describing quantum ballistic
          evolution on some basis $B$ then a step operator $T$ which is
          distinct path generating can be associated with $H$. Details of
          the construction will be given later on.

          As noted earlier a step operator associated with a process is
          defined to be that operator $T$ such that iteration of $T$ (or
          $T^{\dag}$) defines forward (or backward) motion of the process
          on each of the paths.  Note that there is an arbitrariness which
          is decided by convention: namely, that $T$ and not $T^{\dag}$ is
          associated with the forward direction for the process on each
          path.  Also $T$ is defined to be a step operator if there exists
          some basis $B$ (the step basis) such that for each state $\psi$
          in $B$, $T\psi$ and $T^{\dag}\psi$ are orthogonal to $\psi$.

          For  processes such as computations, the step operator $T$ is a
          sum of time ordered products of noncommuting operators for which
          the time interval is $0$ but the noncommutativity and thus the
          time ordering remains.  Computer programs have this property in
          that they are sums of elementary program elements each of which
          is a product of time ordered spatially local actions.  Because
          there is no finite time interval associated with $T$ it can be
          used directly to construct Hamiltonians such as the Feynman
          Hamiltonian of Eq. \ref{ham}.


          Another desirable feature is that for many models of processes,
$H$ can be separated into kinetic and potential energy parts.  This is the case 
for models of quantum Turing machines which will be modelled as motion of a 
head on a one dimensional space lattice of qubits.  For these models,
          $H=KE+PE$ where $KE=K(2-U-U^{\dag})$ and $PE=K(U-T+U^{\dag}-
          T^{\dag})$.  $U$ denotes free motion along the lattice.  In this case 
$KE$ represents the (symmetrized) lattice equivalent of the second
          derivative $Kd^{2}/dx^{2}$ and $PE$ is the interaction potential 
between the head and lattice systems.  In this form $H$ is seen to be similar 
to that used in the tight binding model (with off-diagonal potentials)  
to describe one dimensional particle motion in solids \cite{Erdos}.  This 
similarity will be exploited in future work.

          From the above it is clear that it is important to be able to
          determine if an operator $T$ is distinct path generating.  This
          appears difficult since it appears necessary to examine the
          properties of all powers of $T$ and its adjoint, or instead
          examine the action of $T$ and its adjoint on all states in the
          basis.  It is  desirable to investigate other properties of
          operators which can be proved equivalent to distinct path
          generating and for which effective decision procedures may exist.
          The next section is concerned with two candidate properties,
          orthogonality preservation and stability.

          \section{Orthogonality Preservation, Stability} 
          \label{ipo}
          As was noted in the introduction the work of Bennett and Landauer
          \cite{Bennett,Landauer}  showed that an irreversible computation
          could be made reversible by addition of history degrees of
          freedom.  The expanded process was reversible in that distinct
          states remained distinct with no overlap as the process evolved.  
          From a quantum mechanical viewpoint an important part of this
          work is the preservation of orthogonality relative to some basis. 
          These considerations suggest the following definition: Let
          $\{\vert p_{i}\rangle \}$ denote a basis set for a finite or
          separable Hilbert space $\cal H$ and $T$ be a bounded linear
          operator over $\cal H$: \\
          \\
          {\bf Definition 2} {\em An operator $T$ is {\em weakly
          orthogonality preserving} in the basis $\{\vert p_{i}\rangle \}$
          if for all $i,j,\; \langle p_{i}\vert p_{j}\rangle =0
          \Longrightarrow \langle Tp_{i}\vert Tp_{j}\rangle =0$.} \\
          \\
          Note that the definition applies to all states including those
          for which either $\vert Tp_{i}\rangle =0$ or $\vert Tp_{j}\rangle
          =0$.  It also says nothing about the value of $\langle
          Tp_{i}\vert Tp_{i}\rangle$ if $\vert Tp_{i}\rangle \neq 0$.

          There are many operators which preserve orthogonality weakly in
          some basis.  This includes all normal operators which preserve
          orthogonality weakly in their eigenfunction (or spectral measure)
          basis.  The two dimensional operator  given by the matrix 
          \[    \frac{1}{\sqrt{2}} \left( \begin{array}{lr}
          0 & 0 \\
          1 & 1 \end{array}  \right) \]
          does not weakly preserve orthogonality in the basis $\vert
          1\rangle , \vert 0\rangle$ as it converts $\vert 1\rangle$ to
          $1/\sqrt{2}\vert 0\rangle$ but leaves $\vert 0\rangle$ unchanged
          (other than normalization).  However, it preserves orthogonality
          weakly in the basis $\vert +\rangle , \vert -\rangle$ where
          $\vert \pm \rangle = 1/\sqrt{2} (\vert 1\rangle \pm \vert
          0\rangle)$.  The projection operator $P_{0}$ on the state $\vert
          0\rangle$ preserves orthogonality weakly in the basis $\vert
          0\rangle, \vert 1\rangle$.  It does not preserve orthogonality
          weakly in the basis $\vert +\rangle, \vert -\rangle$.  This would
          be applicable for instance to binary bits represented by the
          states $\vert +\rangle, \vert -\rangle$.  These simple examples
          show the basis dependence of weak orthogonality preservation in
          quantum mechanics.

          These two operators appear equivalent as far as the definition is
          concerned.  Yet one feels intuitively that there is a difference.
          The relevant difference is seen by considering the adjoints.  The
          adjoint  of the first example preserves orthogonality weakly in a
          different basis, namely. $\vert 1\rangle, \vert 0\rangle$,
          whereas the adjoint of the projection operator preserves
          orthogonality weakly in the same basis. 

          From these and other examples, such as the Turing Machine
          examples studied in \cite{Benioff3}, the relevant distinction is
          between operators for which both $T$ and $T^{\dag}$ preserve
          orthogonality weakly on a common basis and those for which the
          basis is different for $T$ than for $T^{\dag}$.  This suggests
          the following definition:  \\
          \\
          {\bf Definition 3} {\em An operator $T$ is {\em orthogonality
          preserving} if both $T$ and its adjoint $T^{\dag}$ are weakly
          orthogonality preserving on the same basis. } \\
          \\
          Based on this definition, one has the following Theorem: \\
          \\
          {\bf Theorem 1} {\em An operator $T$ and its adjoint are
          orthogonality preserving if and only if $T^{\dag}T$ and
          $TT^{\dag}$ commute.} \\
          \\
          The proof of this theorem, which is straightforward, is given in
          Appendix A.

          It is an immediate consequence of this theorem that there is a
          common spectral measure which is a common refinement of those for
          $T^{\dag}T$ and $TT^{\dag}$. By the spectral theorem
          \cite{Halmos1} spectral measures exist for these two operators as
          they are both selfadjoint.  By the above theorem a common
          refinement exists for $T^{\dag}T$ and $TT^{\dag}$. If physicists
          license of usage is allowed, the theorem guarantees the existence
          of an eigenfunction expansion which is common to both $T^{\dag}T$
          and $TT^{\dag}$.  The eigenfunctions are the basis set referred
          to in the theorem.

          It also follows from this that if any of the spectral subspaces
          in the common refinement has dimension $n$ with $n>1$, there
          exists an uncountable infinity of inequivalent bases for which
          $T$ is orthogonality preserving.  To see this take any basis in
          the subspace and change the basis using any unitary operator in
          $U(n)$.  $T$ will also be orthogonality preserving for the new
          basis.

          For the purposes of this paper the requirement that a model
          operator $T$ be orthogonality preserving for a common basis is
          necessary but not sufficient.  To see this recall that  the
          interest here is in constructing Hamiltonians whose time
          evolution gives states representing evolution along trajectories
          of successive steps of the process.  Since $1$ step of the
          process is represented by the model operator $T$, $n$ steps are
          represented by $T^{n}$.  If the Hamiltonian is to properly
          represent the process evolution, then it is necessary that all
          positive powers of $T$ and $T^{\dag}$ be orthogonality
          preserving.  

          This requirement is still not sufficient because, by the
          definition of orthogonal preservation, it means that for each
          $n$, there is a basis set $\{\vert
          p_{i}^{n}\rangle:i=0,1,\cdots\}$ which preserves orthogonality
          for $T^{n}$ and $(T^{\dag})^{n}$.  However, the basis set which
          satisfies the definition may depend on $n$.   

          To avoid this dependence it is required here that there exist a
          common basis for which orthogonality is preserved by all powers
          of $T$ and its adjoint.  More precisely:  \\
          \\
          {\bf Definition 4} {\em An operator $T$ and its adjoint are {\em
          completely orthogonality preserving} if there exists a basis set
          $\{\vert p_{i}\rangle:i=0,1,\cdots\}$  such that for all $i,j$ $
          \langle p_{i}\vert p_{j}\rangle =0 \Longrightarrow  \langle
          T^{n}p_{i}\vert T^{n}p_{j}\rangle =0$ and $\langle
          (T^{\dag})^{n}p_{i}\vert (T^{\dag})^{n}p_{j}\rangle =0$ for
          $n=1,2,\cdots$.} \\
          \\  
          It is clear from the definition that complete orthogonality
          preservation implies orthogonality preservation, but not the
          converse.  Existence of specific examples which are orthogonality
          preserving but not completely orthogonality preserving follow
          from the results in the next sections.  

          Note that for any operator $T$, the set of discrete
          eigenfunctions (if any) of $T$ are completely orthogonality
          preserving on the subspace spanned by the eigenfunctions. This
          basis is not of interest here as it is stationary with respect to
          iterations of $T$.  It is also easy to see that all normal
          operators  ($T$ is normal if $T^{\dag}T = TT^{\dag}$) are
          completely orthogonality preserving.  However, the main interest
          here is in operators $T$ which are not normal.

          It is easy to show by means of specific examples that if an
          operator $T$ is completely orthogonality preserving in a basis
          $B$, this does not imply that for states $\vert p_{i} \rangle$ in
          $B$, that the states $T^{n}\vert p_{i}\rangle$ remain in $B$. 
          This is the case for most unitary operators which are completely
          orthogonality preserving in all bases.  To avoid this the
          additional requirement that a step operator  be stable in an
          orthogonality preserving basis will be used. \\
          \\
          {\bf Definition 5} {\em $T$ and and $T^{\dag}$ are {\em stable
          for some basis} if there exists a basis $B$ such that for all
          $\vert p_{i}\rangle$ in $B$, if $T\vert p_{i}\rangle \neq 0$,
          then  $T\vert p_{i}\rangle =\alpha_{k,i}\vert p_{k} \rangle$ for
          some $\vert p_{k}\rangle$ in $B$.  $\alpha_{k,i}$ is a constant
          $\neq 0$.  A similar statement holds for $T^{\dag}$ for the basis
          $B$.} \\
          \\
          In other words $T$ and its adjoint are stable for some basis $B$
          if $T$ and $T^{\dag}$ map some (or all) states of $B$ into states
          which, except for normalization,  are states in $B$ and
          annihilate the others.  In particular $T$ and its adjoint do not
          map states of $B$ into linear sums of states in $B$.

          The utility of stability is shown by the next two theorems. \\
          \\
          {\bf Theorem 2} {\em An operator $T$ (and its adjoint) are
          orthogonality preserving and stable for some basis $B$ if and
          only if $T$ (and its adjoint) are completely orthogonality
          preserving and stable in $B$.} \\
          \\
          The proof depends on the fact that if $T$ and its adjoint are
          orthogonality preserving and stable on some basis $B$, then one
          can use an inductive argument to show that all powers of $T$ and
          $T^{\dag}$ are orthogonality preserving on $B$, which is
          equivalent to complete orthogonality preservation and stability
          on $B$.  The proof in the other direction is immediate. \\
          \\
          {\bf Theorem 3} {\em A partial isometry $T$ is orthogonality
          preserving and stable in some basis $B$ if and only if $T$ is
          distinct path generating in $B$.} \\
          \\
          The proof of this theorem is given in Appendix A.  

          It follows from these theorems that  if $T$ is required to be
          stable on some basis, then orthogonality preservation and
          complete orthogonality preservation on the same basis are
          equivalent. This raises the question  of the need for complete
          orthogonality preservation, since it appears superfluous.  For
          the purposes of this paper, this question can be explored by
          asking how far can one go, assuming that an operator is
          completely orthogonality preserving without using the assumption
          of stability?   The answer is, "a long way".  This will become
          clear in the following sections.  It will also be seen that the
          eigenfunctions and eigenvalues have a canonical form for the
          Hamiltonian of Eq. \ref{ham} where $T$ is a partial isometry,
          which is stable and orthogonality preserving in some basis.

          From now on the operator $T$ will be limited to be a partial
          isometry.  Recall that $T$ is a partial isometry if and only if
          $T^{\dag}$ is. Model operators for many processes can be
          constructed which are partial isometries.  This includes models
          of quantum computers as (deterministic or nondeterministic)
          Turing machines (Section \ref{qctm}). 

          \section{Power Partial Isometries}
          \label{ppi}
          At this point it is necessary to introduce power partial
          isometries.  A partial isometry $T$ is a power partial isometry
          if all positive powers of $T$ and  its adjoint are partial
          isometries. 

          Power partial isometries (PPI)s were first described by Halmos
          and Wallen \cite{Halmos-wallen} and further developed by others
          \cite{Plebanski,Hoover-lambert,Embry-etal,Herrero}. Related work
          on partial isometries and semigroups of partial isometries
          includes that of  \cite{Plebanski,Erdleyi,Hellwig}. Halmos and
          Wallen have given the main properties of PPIs and proved a
          structure or decomposition theorem.  The relevant mathematical
          results are summarized here.  For details the literature should
          be consulted. 
          For any partial isometry $T$ the projection operators
          $I=T^{\dag}T$ and $F=TT^{\dag}$ define the respective domain and
          range spaces for $T$ and $T^{\dag}$.  That is, $T=TI=FT$ and
          $T^{\dag}=T^{\dag}F=IT^{\dag}$.  

          Let $W$ and $V$ be two partial isometries.  The product $WV$ is a
          partial isometry if and only if \cite{Halmos-wallen} $VV^{\dag}$
          commutes with $W^{\dag}W$.  This will be referred to as the "H-W
          lemma" in Appendix A.

          There are many partial isometries that are not power partial
          isometries.  Halmos and Wallen \cite{Halmos-wallen} have given a
          method of explicit construction of an operator $U$ such that the
          distribution of values of $n$ for which $U^{n}$ is or is not a
          partial isometry is arbitrary.  Their construction is as follows: 
          Let $U_{1}$ denote any contraction operator which is not a
          partial isometry, for example $U_{1}= a(\sigma_{x}-i\sigma_{y})$
          where the $\sigma s$ are the Pauli spin operators and $\vert
          a\vert <1/2$.  For any operator $T$ define the operator
          $D_{T}=(1-TT^{\dag})^{1/2}$ (positive square root implied) and
          the matrix operator $M(A)$ by
          \[ M(A) = \left( \begin{array}{cc}
          A & D_{A} \\
          0 & 0
          \end{array}  \right) \]
          For each $n$ the operator $U_{n}$ is defined inductively for
          $n=2,3,\cdots$ by $U_{n}=M(U_{n-1})$.  For any $k$ where $1 \leq
          k<n$, $U_{n}^{k}$ is a partial isometry, $U_{n}^{n}$ is not, and
          $U_{n}^{n+1}=0$. 

          Let $s$ be any infinite sequence of $0's$ and $1's$.  The desired
          operator is defined by $U=\oplus_{n=1}^{\infty}U_{n}\delta
          _{1,s(n)}$.  Here $\oplus$ denotes the direct sum and $\delta$
          the Kronecker delta.  This result is quite remarkable and has the
          consequence that one has to be very careful to avoid making
          unwarranted assumptions about operators modelling processes such
          as quantum computations.  

          Let $T$ be a power partial isometry.  For each $n=0,1,2,\cdots$
          define the operators $I_{n}$ and $F_{n}$ by
          \begin{eqnarray}
          I_{n} & = & (T^{\dag})^{n}T^{n}  \label{In} \\
          F_{n} & = & T^{n}(T^{\dag})^{n}. \label{Fn}
          \end{eqnarray}
          Since $T$ is a PPI, all the $I_{n}$ and $F_{n}$ are projection
          operators.  The $I_{n}$ and $F_{n}$ form nonincreasing sequences. 
          That is, for all positive n $I_{n}\geq I_{n+1}$ and $F_{n}\geq
          F_{n+1}$.  Also all the $I's$ and all the $F's$ commute among
          themselves and with each other.  That is, for all nonnegative
          $m,n$ $[I_{m},I_{n}]=[F_{m},F_{n}]=[I_{m},F_{n}]=0$. One also has
          that $TI_{n}=I_{n-1}T$ and $TF_{n-1}=F_{n}T$.

          Define $I_{\infty}$ and $F_{\infty}$ to be the respective
          projections on the subspaces $\bigcap_{n=0}^{\infty}I_{n}{\cal
          H}$ and $\bigcap_{n=0}^{\infty}F_{n}{\cal H}$. $I_{\infty}$ and
          $F_{\infty}$ commute with each other and all the $I_{n}$ and
          $F_{n}$. 

          The main property of interest here is the decomposition theorem
          of Halmos and Wallen \cite{Halmos-wallen} which states that every
          power partial isometry has a unique decomposition into a direct
          sum of operators with nonoverlapping domain and range spaces
          (i.e. which reduce $T$) given by 
          \begin{equation}
          T=T_{1}+T_{2}+T_{3} +\sum_{n}T_{4n}.
          \end{equation}
          Here $T_{1}$ is a unitary operator on the range space of
          $I_{\infty}F_{\infty}$, $T_{2}$ is a pure isometry
          (i.e.$T^{\dag}T=1$ on ran$I_{\infty}-I_{\infty}F_{\infty}$,
          $T_{3}$ is a pure coisometry (i.e. $TT^{\dag}=1$) on
          ran$F_{\infty}-I_{\infty}F_{\infty}$, and for each n, $T_{4n}$ is
          a truncated shift of index $n$ on the range space of $P_{n}$.

          A truncated shift of index $n$ is an operator defined on the sum
          of $n$ copies of a Hilbert space which takes any state in the
          $lth$ copy to the same state in the $l+1st$ copy and annihilates
          states in the $nth$ copy.  That is $T_{4n}<\psi_{1} ,\psi_{2},
          \cdots ,\psi_{n-1},\psi_{n}>=<0,\psi_{1}, \psi_{2} , \cdots,
          \psi_{n-1}>$.  For the decomposition theorem the $lth$ copy (for
          $l=1,\cdots ,n$) is the range space of the projection operator
          $P_{n,l}=(F_{l-1}-F_{l})(I_{n-l}-I_{n-l+1})$.  The projection
          operator $P_{n}$ is defined by $P_{n}=\sum_{l=1}^{n}P_{n,l}$.

          A pure isometry is an isometry which is unitarily equivalent to a
          direct sum of copies of the unilateral shift.  It acts like a
          truncated shift of index $\infty$ except that there is no state
          annihilation at any index.  In the above decomposition the domain
          of $T_{2}$ is given by $\sum_{l=1}^{\infty}P_{\infty l} =
          \sum_{l=1}^{\infty}(F_{l-1}-F_{l})I_{\infty}.$  It is easy to see
          that $P_{\infty l}P_{\infty m}=P_{\infty l}\delta_{l,m}$ and
          $T_{2}P_{\infty l}=P_{\infty l+1}$.

          The domain of the pure coisometry (i.e. $T_{3}^{\dag}$ is a pure
          isometry) is given by the projection operator
          $\sum_{l=1}^{\infty}Q_{\infty l} = \sum_{l=1}^{\infty}(I_{n-l}-
          I_{n})F_{\infty}$ where $T_{3}Q_{\infty l}=T_{3}Q_{\infty l-1}$
          and $ T_{3}Q_{\infty 0}=0$.

          In general none or some of the reducing subspaces in the
          decomposition theorem can be empty.  Thus unitary operators, pure
          isometries, pure coisometries, and truncated shifts are all power
          partial isometries.  It is also worth noting that if $T$ is such
          that the unitary componenet is limited to be a sum of bilateral
          shifts and cyclic finite shifts, then $T$ is distinct path
          generating in the more general sense discussed in Section
          \ref{step-qbe}.  

          The following theorem relates complete orthogonality preservation
          to power partial isometries. \\
          \\
          {\bf Theorem 4} {\em Let $T$ be a partial isometry.  Then $T$ and
          $T^{\dag}$ are completely orthogonality preserving if and only if
          $T$ is a power partial isometry.} \\
          \\
          The proof of this theorem is given in Appendix A.

          \section{Quantum Ballistic Evolution and Orthogonality
          Preservation}
          \label{main}
          The material presented so far is sufficient to give a proof of a
          main point of this paper, which is stated in the following
          theorem. The proof is summarized here with details left to the
          reader. \\
          \\
          {\bf Theorem 5} {\em A necessary and sufficient condition that
          there exists a Hamiltonian description of quantum ballistic
          evolution of a process is that there exists a process step
          operator $T$ and a basis $B$ such that $T$ is a partial isometry
          and is stable and orthogonality preserving in $B$.} \\
          \\
          Sufficiency:  Assume $T$ is a partial isometry and is
          orthogonality preserving and stable with respect to a basis $B$.
          By Theorem 3, $T$ and $T^{\dag}$ are distinct path generating.
          The collection of paths can be determined by iteration of $T$ and
          its adjoint on states of $B$.  Use Eq. \ref{ham} to define a
          Hamiltonian $H=K(2-T-T^{\dag})$.

          The time evolution is given by $e^{-iHt}$. A general term in the
          power series expansion of $e^{iKt(T+T^{\dag})}$ where $e^{-iHt}=
          e^{-2iKt}e^{iKt(T+T^{\dag})}$ has the form $\cdots
          (T^{\dag})^{m}T^{n}(T^{\dag})^{k}T^{l}$ where $m,n,k,l$ are
          nonnegative integers. Since $T$ is distinct path generating this
          term describes $l$ steps forward (i.e. the $T$ direction), $k$
          steps backward, $n$ steps forward, $m$ steps backward, etc.,
          along any of the paths generated by $T$ or its adjoint. All of
          these terms in the expansion except the first give $0$ when
          applied to any states of $B$ not in a path.  When applied to any
          state in $B$ in a path, many of the terms give another state of
          $B$ in the path. Since all terms in the expansion are of this
          form, it is clear that $H$ describes quantum ballistic evolution.

          Necessity:  Assume the existence of a Hamiltonian which describes
          quantum ballistic evolution on a basis $B$.  $B$ is clearly not a
          basis of eigenfunctions for $H$. The Hamiltonian can be used to
          construct paths as follows:  For any pair $\vert a\rangle ,\vert
          b\rangle$ of distinct states in $B$ (i.e. $\langle a\vert
          b\rangle =0$), $\vert a\rangle, \vert b\rangle$ are on the same
          path if there exists an $n$ such that $\langle b\vert H^{n}\vert
          a\rangle \neq 0$. $\vert a\rangle,\vert b\rangle$ are not on the
          same path if $\langle b\vert H^{n}\vert a\rangle =0$  for all
          $n$.  Since $H$ describes quantum ballistic evolution, if $\vert
          a\rangle,\vert b\rangle$ with the two states distinct are on the
          same path, there is a least $n$ such that $\langle b\vert
          H^{n}\vert a\rangle \neq 0$.  Denote the least $n$ by $n_{ba}$. 
          Note that $n_{ba}=n_{ab}$. 

          A pair of distinct states $\vert a\rangle,\vert b\rangle$ are
          adjacent if $n_{ab}=1$.  In general a state can have
          $0,1,2,\cdots...$  adjacent states. Since $H$ describes ballistic
          evolution, motion is restricted to distinct paths only.  Thus, at
          most, $2$ states can be adjacent to a given state in $B$. A state
          with $0,1,2$ states adjacent is on no path (or a path of length
          $0$), is a terminal state of a path, or is an interior state of a
          path respectively.  Note also by the definition of quantum
          ballistic evolution, for any pair $\vert a\rangle,\vert b\rangle$
          of distinct states in $B$, $\langle b\vert H\vert a\rangle \neq
          0\Rightarrow \langle b\vert H\vert a\rangle =c$ where $c$ is a
          constant independent of $\vert a\rangle,\vert b\rangle$.

          A step operator can be defined as follows: Choose states of $B$
          until a pair $\vert a\rangle,\vert b\rangle$ of adjacent distinct
          states are found. Set $T\vert a\rangle =\vert b\rangle$ and
          $T^{\dag}\vert b\rangle =\vert a\rangle$. Continue searching $B$
          for states adjacent to and distinct from either $\vert a\rangle$
          or $\vert b\rangle$. If a state $\vert d\rangle$ adjacent to
          $\vert b\rangle$ is found, set $T\vert b\rangle =\vert d\rangle$
          and $T^{\dag}\vert d\rangle =\vert b\rangle$.  If no state
          adjacent to $\vert b\rangle$, other than $\vert a\rangle$ is
          found, set $T\vert b\rangle=0$.  The same construction applies
          for a state, if any, adjacent to $\vert a\rangle$ with $T^{\dag}$
          exchanged for $T$. If no states are both distinct from and
          adjacent to $\vert a\rangle$, set $T\vert a\rangle=0$.

          Continuing in this manner by searching through all states of $B$
          defines all paths of $H$ and an associated step operator $T$
          which is a partial isometry.  Since $T$ is distinct path
          generating, by Theorem 3, $T$ is orthogonality preserving and
          stable on $B$, and the theorem is proved.


          The necessity proof of the theorem has an arbitrariness in the
          choice of directions on each path for $T$ and $T^{\dag}$. This
          can be seen by exchanging $T$ for $T^{\dag}$ on one or more
          paths.  From this one sees that for a given quantum ballistic
          Hamiltonian that describes motion on $n$ distinct paths, there
          are $2^{n}$ possible choices for the associated step operator
          $T$. 

          This arbitrariness is equivalent to the possiblity of 
          construction of wave packets which can move in either of two
          directions on each path.  However if $H$ describes quantum
          ballistic evolution for some process with a defined "forward"
          direction with increasing time, then one of the possible choices
          of $T$ will be the step operator for the process.  Which one is
          chosen will depend on external conditions, such as the choice of
          possible initial states.  This is case for quantum computations,
          including reversible ones, where there are well defined forward
          and backward directions.  The choice of whether $T$ or $T^{\dag}$
          is associated with the forward direction for all the paths is
          chosen by convention.

          It should be noted that on any basis for which $T$ is
          orthogonality preserving, the Hamiltonian generating quantum
          ballistic evolution for $T$ is not orthogonality preserving on
          the basis.  Since Hamiltonians are selfadjoint and thereby
          normal, they are completely orthogonality preserving on some
          basis.  However they are not even orthogonality preserving on any
          quantum ballistic basis for $T$. For example, $T+T^{\dag}$, which
          is in essence the Hamiltonian of Eq. \ref{ham}, is not
          orthogonality preserving on any quantum ballistic basis for $T$. 
          On the other hand, any unitary operator, such as $e^{-iHt}$, is
          completely orthogonality preserving on all basis sets. In fact
          unitary operators are the only ones with this property.

          It is of interest to examine the effect of replacing in Theorem 5
          stability and orthogonality preservation with complete
          orthogonality preservation.  It is clear that necessity still
          holds but sufficiency fails.  However sufficiency almost works at
          least if $T$ is a partial isometry.  To see this note that if $T$
          is a partial isometry and is completely orthogonality preserving,
          it is a power partial isometry and the decomposition theorem
          applies.  All components of the decomposition are distinct path
          generating, except for those components in the unitary part which
          are not equivalent to (copies of) the bilateral shift or to
          cyclic orbits on a basis. For these components sufficiency fails.
          This is the sense in which complete orthogonality preservation
          without the separate assumption of stability "almost works".

          This can be said in another way by noting that if $T$ is a power
          partial isometry, then $H$ defined by Eq. \ref{ham} is partially
          quantum ballistic.  This concept, which will be used later, means
          that on some of the reducing subspaces $H$ is quantum ballistic
          and on others it is not.  If $T$ is a PPI then $H$ is quantum
          ballistic except possibly on components in the unitary part that
          are not equivalent to the bilateral shift or finite cyclic
          shifts.

          It is of interest to examine the eigenfunctions and eigenvalues
          of the Hamiltonian given by Eq. \ref{ham} in the case that $T$
          satisfies the conditions of Theorem 5. In this case $T$ is a
          power partial isometry which models some process which evolves
          quantum ballistically on $B$.  For any such $T$, the
          eigenfunctions and eigenvalues of the Hamiltonian given by Eq.
          \ref{ham} all have the same form.  

          To see this it is sufficient to consider the unitary, isometric,
          coisometric, and truncated shift components separately.  Paths in
          each part are defined by a basis set $\{\vert n,l\rangle \}$
          where $T\vert n,l\rangle =\vert n+1,l\rangle $ and $T^{\dag}\vert
          n,l\rangle =\vert n-1,l\rangle $.  The label $l$ stands for the
          fact that $T$ may be a direct sum of PPIs and for the fact that
          the subspaces in the direct sum for the truncated shift can be
          multidimensional.\footnote{In mathematical language the basis
          ranging over all $l$ for a fixed value of $n$ span a wandering
          subspace for the part being considered.}  The basis state labels 
          are kept simple for the sake of clarity.  

          Eigenfunctions and eigenvalues for $H$ can be determined by 
          writing the eigenfunction in the general form \cite{Merzbacher}
          (the $l$ label is suppressed):
          \begin{equation} 
          \Psi_{k} = \sum_{n=-L}^{M}(Ae^{ikn}+Be^{-ikn})\vert n\rangle
          \label{nAB}
          \end{equation}
          where $T\vert M\rangle = T^{\dag}\vert -L\rangle =0$.  Here $k$
          denotes a momentum. The values of $M,N$ depend on the part being
          considered.  Eigenvalues and eigenfunctions are obtained by
          writing $(E-H)\Psi_{k}=0$, equating coefficients of each basis
          state to $0$, and solving the system of linear equations so
          obtained. For all parts eigenvalues are given by 
          \begin{equation}
          E_{k} = 2K(1-\cos k) \label{eigenvalue}
          \end{equation}
          where $k$ ranges from $-\pi$ to $\pi$. 

          For the unitary part there are two type of paths to consider, two
          way infinite and finite cycles or orbits.  Eigenfunctions are
          obtained from Eq. \ref{nAB} by setting $B=0$.  For the infinite
          paths, $M=L=\infty$ and
          \begin{equation}
          \Psi_{k} = \frac{1}{\sqrt{2\pi}}\sum_{n=-
          \infty}^{\infty}e^{ikn}\vert n\rangle \label{u-eigen}
          \end{equation}
          The momentum $k$ can assume all values.  For the orbits, $L=0$
          and $M$ is finite with $T\vert n,l\rangle = \vert n+1,l \rangle$
          if $n<M$ and $T\vert M,l\rangle = \vert 0,l\rangle$.  $T^{\dag}$
          moves along the orbit in the opposite direction.  The
          eigenfunctions, which are also eigenfunctions of $T$ and
          $T^{\dag}$ separately, are given by
          \begin{equation}
          \Psi_{k} = \frac{1}{\sqrt{M+1}}\sum_{n=0}^{M}e^{ikn}\vert
          n\rangle \label{cycle}
          \end{equation}
          where $k$ is discrete with $k=2\pi m/(M+1)$ with $m=1,2,\cdots
          ,M+1$.

          For the pure isometric part which is a direct sum of copies of
          the unilateral shift,  $L=0,M=\infty$, and the eigenfunctions
          have the form (the index $l$ is suppressed)
          \begin{equation}
          \Psi_{k,\geq 0} = C'\sum_{n=0}^{\infty}\sin k(n+1)\vert
          n\rangle. \label{geq0}
          \end{equation}
          For the coisometric part $L=-\infty ,M=0$  and 
          \begin{equation}
          \Psi_{k,\leq 0} = C'\sum_{n=-\infty}^{0}\sin k(n-1)\vert
          n\rangle. \label{leq0}
          \end{equation}

          For both these cases $k$ can assume any value between $-\pi$ and
          $\pi$. C' is a normalization constant.  The isometric and
          coisometric eigenfunctions represent standing waves for complete
          reflection from a barrier at state $\vert 0\rangle $ which
          defines the beginning or terminus (for $T$) of a path.

          For a truncated shift of index $N,\; L=0,\; M=N-1$ and the
          eigenstates are given by
          \begin{equation}
          \Psi_{k}=\frac{1}{(N)^{1/2}}\sum_{n=0}^{N-1}\sin k(n-N) \vert
          n\rangle  \label{eq:b-eigens}
          \end{equation}
          The eigenvalues, given by  Eq. \ref{eigenvalue}, are discrete
          with $k=m\pi /(N+1)$ and $m$ takes on integral values from $1$ to
          $N$.  The value $m=0$ is excluded as the eigenfunction is
          identically $0$ for this case. The eigenfunctions are $0$ at $n=-
          1$ and $n=N+1$ and correspond to bound states in a square well
          with completely reflecting walls at $n=-1$ and $n=N+1$.

          The above shows that $H$ given by Eq. \ref{ham} has the
          eigenvalues and eigenfunctions corresponding to quantum ballistic
          evolution on basis state paths.  The types of paths correspond to
          the types of shifts in the decomposition of $T$.  Furthermore any
          wave packet state along a path defined by Eq. \ref{packet} moves
          along the path spreading out as it moves.  

          The reason for presenting a description of eigenvalues and
          eigenfunctions and wave packet spreading is that it gives a
          complete description of eigenvalue and eigenfunction structure
          for all processes modelled in quantum mechanics by step operators
          $T$ which are partial isometries and are stable and orthogonality
          preserving on some basis.  The differences in individual
          processes show up in the multiplicities and in the complexity and
          description of basis states in the paths.  Examples which
          illustrate this are given next.

          \section{Examples} 
          \label{examples}
          Some simple examples will be considered to illustrate aspects of
          the preceding.  The physical model considered will be that
          appropriate for discussion of Turing machines. To this end the
          model consists of a two way infinite one dimensional lattice of
          qubits represented as spin $1/2$ systems fixed at the lattice
          points.  A head moves along the lattice interacting locally with
          spins.  For the simple examples considered the head is spinless. 
          For Turing machines, the head has a large but finite number of
          internal states, such as the $2L+1$ spin projection states for a
          particle with spin $L$.

          The above description gives an uncountable infinity of lattice
          spin basis states which can be written in the form $ \vert
          s\rangle =\otimes_{j=-\infty}^{\infty}\vert s(j)\rangle$ where
          $s$ is any function $s:Z\rightarrow \{0,1\}$. $Z$ is the set of
          integers.  In order to work with a separable Hilbert space $s$ is
          limited to be any function with at most a finite number of values
          different from $0$.  The resulting Hilbert space ${\cal H}_{00}$
          for the lattice spins is spanned by all spin states with tails of
          $0$ in both directions.

          The  overall model (separable) Hilbert space is spanned by
          vectors of the form $\vert l,j,s\rangle $ where $l$ is the
          internal head state, $j$ is the head lattice position and $\vert
          s\rangle$ is limited to $0$ tail states.  Here each of the
          lattice systems and the head are taken to be distinguishable
          particles to avoid the complications of antisymmetrization which
          are not relevant here.  Also the direction of quantization is
          taken to be along the $z$-axis for each of the spins.  

          \subsection{Motion in the Presence of 0s}
          \label{0s-motion}
          In this example the (spinless) head moves along the lattice only
          at spin $0$ lattice sites.  The operator $T$ and its adjoint are
          defined by 
          \begin{eqnarray}
          T & = & \sum_{j=-\infty}^{\infty}P_{0,j}UP_{j}  \label{0motion}
          \\
          T^{\dag} & = & \sum_{j=-\infty}^{\infty}P_{0,j}P_{j}U^{\dag}.
          \label{0motion1}
          \end{eqnarray}
          Here $P_{i,j}$ is the projection operator for finding the site
          $j$ lattice spin in state $\vert i\rangle$ with $i=0$ or $i=1$,
          $P_{j}$ is the projection operator for the head at site $j$, and
          $U$ is the unitary operator shifting the head one site to the
          right $(UP_{j}=P_{j+1}U)$.  The sum is over all lattice sites.
          The lattice spin projectors commsum is over all lattice sites.

          It is easy to verify that $T$ is a power partial isometry.  In
          particular $I_{n}$ and $F_{n}$, Eqs. \ref{In} and \ref{Fn}, given
          by 
          \begin{eqnarray}
          I_{n} & = & \sum_{j=-\infty}^{\infty}P_{0,j+n-1}P_{0,j+n-2}\cdots
          P_{0,j} P_{j} \\
          F_{n} & = & \sum_{j=-\infty}^{\infty}P_{0,j+n-1}P_{0,j+n-2}\cdots
          P_{0,j} P_{j+n}
          \end{eqnarray}
          are projection operators.   Also the $I_{n}$ and $F_{m}$ commute
          among themselves and with each other.

          It follows from Theorem 4 that $T$ and $T^{\dag}$ are completely
          orthogonality preserving. The common basis set is the set of all
          $\vert j,s\rangle$ as defined above.  It is also clear by
          inspection of the definition of $T$ and its adjoint that $T$ and
          $T^{\dag}$ are stable in this basis.  $T$ also has pieces in each
          of the components given by the decomposition theorem.  

          The component subspaces are defined by properties of the spin
          lattice.  The unitary part of $T$ acts in the subspace which is 
          spanned by $\vert j,s\rangle$ where $s$ is the constant $0$
          sequence.  On this space $T$ is the bilateral shift.  

          The other component subspaces are characterized by those
          sequences $s$ which contain a finite positive number of $1's$. 
          Let $m_{u}$ and $m_{l}$ be the greatest and least integer
          respectively such that $s(m_{u}) =s(m_{l}) = 1$.  Then $T$
          restricted to the subspace spanned by $\{\vert j,s\rangle :j\geq
          m_{u}+1$ is an isometry in that it is a unilateral shift.  Note
          that $T^{\dag} \vert m_{u}+1,s\rangle = 0$.  Similarly, $T$
          restricted to the subspace spanned by $\{\vert j,s\rangle :j\leq
          m_{l}$ is a coisometry in that $T^{\dag}$ is a unilateral shift. 
          Note that $T \vert m_{l},s\rangle = 0$.

          Those $s$ with one solid block of $1's$ surrounded on both sides
          by $0's$ extending to infinity, are included in the pure
          isometric and pure coisometric subspaces of $T$ only. Those $s$
          with one or more groups of $0's$ separated by $1's$ on both
          sides, T also has truncated shift components in addition to the
          terminal isometric and coisometric components.  For example let
          $s$ be such that all lattice spins are down except at locations
          $N$, and $N+W+1$.  Figure 2 shows details.  Here $T$ is a
          truncated shift of index $W+1$ on the subspace spanned by
          $\{\vert j,s\rangle :N+1\leq j\leq N+W+1\}$. 

          The eigenfunctions and eigenvalues for the Hamiltonian of Eq.
          \ref{ham} are included in the results of the previous section. 
          The Hamiltonian has a rich eigenfunction structure in that it
          corresponds to a collection of many different Hamiltonians, one
          for each lattice state $\vert s\rangle$, each describing head
          motion on a one dimensional lattice in the potential environment
          given by $\vert s\rangle$.

          For the unitary component the head eigenfunctions, $\psi_{k}$,
          are given by Eq. \ref{u-eigen}.  Eigenvalues are given by Eq.
          \ref{eigenvalue} with all values of the momentum $k$ between $-
          \pi$ and $\pi$ allowed.  For the example shown in Figure 2, the
          isometric and coisiometric eigenfunctions describe respectively
          righthand standing waves reflecting off the $1$ at site $N+W+1$
          and lefthand standing waves reflecting off the site at location
          $N+1$ .  These are given by 
          \begin{equation}
          \psi_{k} = C'\sum_{n=N+W+1}^{\infty}\sin (k(n-N-W-1))\vert
          n,\rangle. \label{sc>N+M}
          \end{equation}
          for the righthand state and
          \begin{equation}
          \psi_{k} = C'\sum_{n=-\infty}^{N+1}\sin (k(n-N-1))\vert
          n\rangle \label{sc<-N}
          \end{equation}
          for the lefthand head state.  These equations are obtained from
          Eqs. \ref{geq0} and \ref{leq0}.

          There is an assymetry in the barrier locations in that for the
          isometry the barrier is at the position of the $1$ whereas for
          the coisometry it is displaced one site to the right of the $1$
          position.  This is a consequence of the assymetry implicit in the
          definition of $T$ and its adjoint.  $T$ reads the location from
          which the head moves and $T^{\dag}$ reads the location to which
          the head moves. The same displacement effect for the bound states
          is shown in Fig. 2.

          For the regions of $0's$ in between the $1's$ the bound state
          eigenstates are given by 
          \begin{equation}
          \psi_{k}=\frac{1}{(W+1)^{1/2}}\sum_{n=N}^{N+W+2}\sin k(n-N-W-2)
          \vert
          n\rangle  \label{eq:b-eigens0}
          \end{equation}
          for the region of $0's$ between sites $N$ and $N+W+1$, Fig. 2,
          The eigenvalues are given by  Eq. \ref{eigenvalue} where $k=m\pi
          /(W+2)$ and $m$ takes on integral values from $1$ to $W+1$.  

          These results hold for all values of $W\geq 2$.  For $W=1$ there
          is just one $0$ between the two $1s$.  In this case there are two
          eigenvalues, $k=0$ with $E=K$  and $k=\pi$ with $E=3K$. It is
          also clear that any region of one or more $1's$ acts as a
          completely reflecting barrier with no communication between
          states on either side.  Thus states on both sides of the barier
          are completely independent of one another. 

          The bound state components correspond to infinite square wells in
          the limit of 0 lattice spacing.  To see this let $d$ denote the
          lattice spacing.  From Eq. \ref{ham} one has that $\langle
          H\rangle = E = 2K$ for any head state in spin-up lattice regions.
          Since $K$ is proportional to $d^{-2}$ one has that $E\rightarrow
          \infty$ as $d\rightarrow 0$.  If the potential width is given by
          $D = (W+2)d$, Eq.~\ref{eigenvalue} gives in the limit
          $d\rightarrow 0$ with $K=c/d^{2}$, $E= K(m\pi /(W+2))^{2} =
          c(m\pi /D)^{2}$.  This corresponds to the usual continuum limit
          \cite{Merzbacher}.

          These results extend to more complex expressions on the tape. 
          For example, let $\vert s\rangle$ contain $M$ bands of
          $m_{1},m_{2},\cdots m_{M}$ $0s$ separated by bands of $1s$.  Each
          band of $0s$ is equivalent to a square well with $m_{1}-1, m_{2}-
          1,\cdots m_{M}-1$ eigenstates of the form of Eq.
          \ref{eq:b-eigens0} with eigenvalues given by Eq. \ref{eigenvalue}
          with $W=m_{1},m_{2},\cdots m_{M}$.  Linear superpositions of
          eigenstates from different wells are also eigenstates, but only
          for component eigenstates with equal energy eigenvalues.

          The above shows how the one simple Hamiltonian in this example
          combines with all lattice spin states which are products of
          $\vert 1\rangle $ or $\vert 0\rangle$ spin projection states to
          generate all possible distributions of completely reflecting barriers 
and the lattice equivalent of infinitely high square wells for particle 
motion on a one dimensional lattice.  

          Except for the unitary part, each of these components in the
          decomposition is of infinite multiplicity in that there are an
          infinite number of different $0-1$ spin distributions
          contributing to each component.   Each lattice spin state $\vert
          s\rangle$ with at least one $1$ contributes to the isometric and
          coisometric parts. Each  $\vert s\rangle$ with $n\: 0s$ between
          $1s$ contributes to the truncated shift of index $n$ part.   Thus
          the general definition of path given in Section \ref{step-qbe}
          applies here.  Only one distribution, that with $0s$ only,
          contributes to the unitary part.  

          In many aspects the properties of this first example are obvious
          or straightforward.  No bit transformations are involved.  These
          are introduced in the next example. 

          \subsection{Arbitrary Bit Transformation} 
          \label{Bit-rotation}
          As will be seen in the next section, elementary program elements
          of Turing computations consists of two types: those in which the
          head state changes after one iteration and those in which the
          head state is fixed.  An example of an operator which models the
          latter type is given by 
          \begin{equation}
          T= \sum_{j=-\infty}^{\infty}v_{j}P_{0,j}UP_{j}
          \end{equation}
          Here $v_{j}$ is any unitary operator in $U(2)$ which transforms
          the site $j$ lattice spin state. An example is the "Fourier"
          transformation \cite{BV,Simon} $v_{00}=
          1/\sqrt{2}(\sigma_{x}+\sigma_{z})$ which has been used in quantum
          computation \cite{Simon,DJ}.  

          For Turing machine steps, $v_{j}$ is independent of $j$. However
          the following discussion for this example remains valid if
          $v_{j}$ depends on $j$.  For deterministic computations $v_{j}$
          is either $1$ or $\sigma_{x,j}$.  For nondeterministic
          computations $v_{j}$ is not restricted.  However it has been
          shown by Deutsch \cite{Deutsch} and Bernstein and Vazirani
          \cite{BV} that it is sufficient to limit $v_{j}$ to the
          deterministic operators plus a rotation by an irrational multiple
          of $\pi$.  This limitation will not be used here.

          It is easy to see that $T$ is a power partial isometry and is
          completely orthogonality preserving.  It is quite similar to the
          previous example in that all parts of the Halmos-Wallen
          decomposition are present except the unitary and coisometric
          parts.  For each component of the decomposition which is present,
          eigenfunctions and eigenvalues have exactly the same form and
          values as those for the previous example.  

          The main difference is in the stable basis to which they refer. 
          Here the basis consists of states of the form $\vert
          j,S_{j}\rangle$ where
          \begin{equation}
          \vert j, S_{j}\rangle = \vert j\rangle \otimes \vert
          S_{<j}^{M}\rangle \otimes \vert S_{\geq j}\rangle \label{jSj} 
          \end{equation}
          where $j$ denotes the head lattice position and $M$ is any integer 
$\leq j$.   

          For all nonempty parts in the Halmos-Wallen decomposition $\vert
          S_{<j}^{M}\rangle$ is given by
          \begin{equation}
          \vert S_{<j}^{M}\rangle = \otimes _{k=M}^{j-1} v_{k}\vert
          0\rangle_{k} \otimes v_{M-1}\vert 1\rangle_{M-1} \otimes
          \psi_{<M-1} \label{S<l}
          \end{equation} 
          where $M\leq j$.  The state $\psi_{<M-1}$ denotes an arbitrary
          state in the spin space for lattice spins at positions $<M-1$ and
          represents the fact that any basis set over this lattice spin
          subspace consistent with the tail condition is allowed. The
          dimension of this subspace is countably infinite.  The
          arbitrariness is possible because no state in the basis of Eq. 
\ref{jSj} has the head in this region.  Orthogonality is guaranteed because 
for a fixed $j$ and $L\neq M$, 
$\langle S_{<j}^{L}\vert S_{<j}^{M}\rangle =0$ independent of the tail states.  
This arises because the factor $\langle 1\vert v^{\dag}v\vert 0\rangle=0 $ at 
sites $L$ or $M$ if $M<L$ or $L<M$.

          For the isometric part $\vert S_{\geq j}\rangle =\otimes_{k}
          \vert 0\rangle_{k\geq j}$.  For the truncated shifts, 
          \begin{equation}
          \vert S_{\geq j}\rangle = \otimes_{k=j}^{N-1} \vert 0\rangle_{k}
          \otimes \vert 1\rangle_{N} \otimes \psi_{>N} \label{Sgeql}
          \end{equation}
          where $N\geq j$.  The argument given above for $\psi_{<M-1}$
          applies to $\psi_{>N}$ with $T$ replacing $T^{\dag}$.  

The reason there are no unitary or coisometric parts is that the state 
$S_{<j}^{M}$ with $M=-\infty$ is orthogonal to any state consistent with the 
$0$ tail condition.  Cyclic orbits cannot occur because of the different bases 
used for the bits to the right and to the left of the head position.

          It is left to the reader to see that $T$ is stable and
          orthogonality preserving with respect to this basis. Theorem 5
          gives the result that the Hamiltonian of Eq. \ref{ham}
          \cite{Feynman} describes quantum ballistic evolution on this
          complex basis. For this and the previous examples, arguments
          given earlier show that complete orthogonality preservation is
          equivalent to orthogonality preservation and stability on the
          described basis.  

          These considerations emphasize how orthogonality preservation and
          stability depend on the basis set chosen.  For the first example,
          $T=\sum_{j}P_{0,j}UP_{j}$ is orthogonality preserving and stable
          in the $0,1$ computation basis.  However it is not orthogonality
          preserving in the $v\vert 0\rangle, v\vert 1\rangle$  basis since
          for most $v$,  $\langle 1\vert v^{\dag}P_{0}v\vert 0\rangle \neq
          0$.

          For this example, eigenfunctions and eigenvalues can be easily
          found for the Hamiltonian of Eq. \ref{ham}.   It is left to the 
          reader to see that they are the same as those for the isometric
          and truncated shift parts of the previous example.  The main
          difference between this example and the previous one is the
          complexity of the basis for which $T$ is orthogonality
          preserving. In this basis the head motion becomes entangled with
          changes in the spin projections on the lattice.  Also the lack of
          unitary and coisometric components is a result of the $0$ tail
          condition.

          \section{Quantum Computers: Turing Machines}
          \label{qctm}
          Step operators for quantum Turing machines (QTM)s can be defined
          based on the physical model given earlier.  To this end let
          $f,d,v$ be three functions with a common domain $D\subseteq 
          [0,N]\times [0,1]$ and respective ranges in $[0,N], [1,\dag ],
          U(2)$.  Here   $[0,N]$ is a finite set of whole numbers from $0$
          to $N$ representing the spin states of the head, $[0,1]$ denotes
          the two states of each qubit or lattice spin in whatever basis is
          chosen as the computation basis, and $U(2)$ denotes the set of 2
          dimensional unitary operators.  Each quantum Turing machine is
          represented by a triple $f,d,v$ of such functions with the model
          operator $T^{f,d,v}$ given by
          \begin{equation}
          T^{f,d,v} = \sum_{l,s\epsilon D} T^{f,d,v}_{ls} \label{Tturing}
          \end{equation}
          as a finite sum over program element operators.  $l$ and $s$
          denote elements of $[0,N],[0,1]$.  From now on the superscript
          $"f,d,v"$ will be suppressed.

          The program element operators have the form
          \begin{equation}
          T_{ls}=\sum_{j=-\infty}^{\infty}u^{f(l,s)}Q_{0}(u^{\dag})^{l}
          v_{lsj}P_{sj}U^{d(ls)}P_{j} \label{Tls}
          \end{equation}
          where $Q_{l}$ is the projection operator for finding the head in
          state $\vert l\rangle$, $u$ is the unitary operator which shifts
          the head state up by one unit, $uQ_{l}=Q_{l+1}u \bmod N$, and
          $v_{lsj}$ is a unitary operator which changes the state of the
          site $j$ lattice spin.  The action of $v_{lsj}$ on the site $j$
          spin is the same for all values of $j$.  The other operators are
          as previously defined. 

          This definition uses the work of Bernstein and Vazirani \cite{BV}
          which shows that any QTM with program elements with no head
          motion can be replaced by an equivalent machine with program
          elements in which the head moves either one cell to the left or
          to the right.

          Both deterministic and nondeterministic QTMs are included.  In
          the usual basis with lattice spins up or down, a deterministic
          quantum Turing computation is one for which the spin change
          operators $v_{lsj}$ in each of the program elements (Eq.
          \ref{Tls}) in the sum over $l,s$ (Eq. \ref{Tturing}) are
          restricted to be either the identity or the spin flip operator
          $\sigma_{x}$.  By use of suitable unitary transformations this
          definition can be applied to any lattice spin bit basis chosen as
          the computation basis.

          The program element operators of Eq. \ref{Tls} are of two types
          depending on whether $f(l,s)\neq l$ or $f(l,s)= l$. In the first
          case it is easy to verify that $T_{ls}^{2}=(T_{ls}^{\dag})^{2}=0$
          and $T_{ls}$ is a partial isometry.   For the case where $f(l,s)=
          l$  a straightforward calculation gives (Eqs. \ref{In} and
          \ref{Fn})
          \begin{eqnarray}
          I_{lsn} & = & Q_{l}\sum_{j=-\infty}^{\infty}\prod_{h=0}^{n-
          1}P_{s,j\pm (h)}P_{j} \\
          F_{lsn} & = & Q_{l}\sum_{j=-\infty}^{\infty}\prod_{h=0}^{n-
          1}v_{lsj\pm h}P_{s,j\pm h}v^{\dag}_{lsj\pm h}P_{j\pm n}
          \end{eqnarray}
          Here $"\pm"$ denotes $+$ if $d(l,s)=1$ and $-$ if $d(l,s)=\dag$. 

          It is straightforward to show that for all $m,n$ $I_{lsn}$ and
          $F_{lsm}$ are projection operators and that
          $[I_{lsn},F_{lsm}]=0$.  So all $T_{ls}$ as defined by Eq.
          \ref{Tls} are power partial isometries.

          This result, although of interest, is not sufficient since one is
          interested in the overall computation process operator $T$, not
          just the program elements.  Iteration of $T$ leads to interaction
          among the different program elements.  In the interest of
          simplicity it is assumed here any history recording steps can be
          added by additional steps on a $1$-tape machine.  In this fashion
          $T$ with or without any history has the form of Eq. \ref{Tturing}
          with all individual elements given by Eq. \ref{Tls}.   This
          requirement that $T$ be a one tape machine is not essential, as
          the arguments can be extended to apply to machines with more than
          one tape. 

          The definition of $T$ given above for quantum Turing machines is
          quite general.  For example step operators for irreversible
          Turing computers are included as are many $T$ which are not
          partial isometries or do not describe ballistic evolution.  It is
          thus of interest to relate the necessary and sufficient
          conditions for quantum ballistic evolution, given in the previous
          sections, to detailed properties of QTM step operators as defined
          by Eqs. \ref{Tturing} and \ref{Tls} and used in the Hamiltonian
          of Eq. \ref{ham}. 

          By definition $T$ is a partial isometry if and only if
          $I_{1},F_{1}$, defined by Eqs. \ref{In} and \ref{Fn}, are
          projection operators.  That is they satisfy $I_{1}^{2}=I_{1}$ and
          $F_{1}^{2}=F_{1}$.  $I_{1},F_{1}$ are given by
          \begin{eqnarray}
          I_{1} & = & \sum_{(ls),(l's')\epsilon D}T^{\dag}_{ls}T_{l's'} \\
          F_{1} & = & \sum_{ls\epsilon D}T_{ls}T_{ls}^{\dag} \\
          \mbox{} & = & \sum_{ls\epsilon D} F_{ls1}
          \end{eqnarray}
          From the definition of $T$ one sees that all nondiagonal elements
          vanish in the definition of $F_{1}$.

          A straightforward calculation using the above shows that 
          \begin{equation}
          I_{1}^{2} =\sum _{(ls),(l's'),(mt)} T^{\dag}_{ls}F_{l's'1}T_{mt}
          =I_{1}+ \sum_{(ls)(mt)} \sum_{(l's')\neq
          (ls),(mt)}T^{\dag}_{ls}F_{l's'1}T_{mt}. \label{I2}
          \end{equation}
          So $I_{1}^{2}$ is a projection operator if and only if the 
          righthand double sum in Eq. \ref{I2} equals $0$.  Carrying out a similar
          calculation for $F_{1}$ shows that $F_{1}$ is a projection
          operator if and only if 
          \begin{equation}
          \sum_{(ls)\neq (l's')}F_{ls}F_{l's'} =0. \label{F2}
          \end{equation}
          Although these conditions are necessary and sufficient for
          determining if $T$ is a partial isometry, they are abstract.  It
          would be good to have more concrete conditions related to the
          detailed properties of the computation.  To this end note that a
          sufficient condition for $T$ to be a partial isometry is that
          $I_{1}=\sum_{ls}I_{ls1}$, that is, all nondiagonal terms
          $T_{ls}^{\dag}T_{mt}$ with $(ls) \neq (mt)$ equal $0$.  (An
          equivalent expression of this is that the terms in the sum of Eq.
          \ref{Tturing} are pairwise orthogonal on the left.) 

          This condition is the quantum mechanical equivalent of the
          classical requirement that no pair of program elements takes two
          different computation states into the same state.  That is, in
          the reverse computation at most one elementary step is active at
          each stage. The already existing condition that nondiagonal terms
          equal $0$ in $F_{1}$  ensure that in the forward computation at
          most one elementary step is active at each stage. In quantum
          mechanics where nondeterminism and quantum parallelism occur,
          more than one elementary term $T_{ls}$ can be active in a stage.

          A sufficient condition for the validity of $I_{1}=
          \sum_{ls}I_{ls1}$ is that the function $f$ in Eq. \ref{Tturing}
          be $1-1$ on $D$.  A less restrictive condition is the following:  For
          all $(ls),(mt)$ in $D$ if $(ls)\neq (mt)$ and $ f(ls)=f(mt)$,
          then $d(ls)=d(mt)$ and for each $j$, $P_{sj}v_{lsj}^{\dag}v_{mtj} 
P_{tj}=0$. Here   $P_{s},P_{t}$ are projection operators for single lattice spin
          states $s,t$.  This condition, called here condition X, follows from 
the properties of the factors in 
          \begin{equation}
          T^{\dag}_{ls}T_{mt}= u^{l}Q_{0}u^{f(ls)-
          f(mt)}Q_{0}(u^{\dag})^{m}\sum_{j,k}P_{sj}v^{\dag}_{lsj}v_{mtk}
          P_{tk}P_{j}(U^{\dag})^{d(ls)}U^{d(mt)}P_{k}
          \end{equation}

          From this equation it can be seen that Condition X is equivalent
          to the condition that the nondiagonal terms $T^{\dag}_{ls}T_{mt}
          =0$ individually.  It is thus a sufficient but not necessary
          condition for $T$ to be a partial isometry.

          It remains to show that the requirement of quantum ballistic
          evolution is not empty for nondeterministic QTMs.  It is first
          shown that there exist step operators for nondeterminstic QTMs
          which are partially quantum ballistic.  That is, they are quantum
          ballistic on some subspaces but not on other subspaces.  To this
          end consider the following example step operator and its adjoint:
          \begin{eqnarray}
          T & = & Q_{0}\sum_{j} v_{00j}P_{0j}UP_{j}
          +uQ_{0}\sum_{j}P_{1j}U^{\dag}P_{j}+Q_{1}\sum_{j}\sigma_{xj}P_{1j}
          U^{\dag}P_{j} \label{Tex1} \\
          T^{\dag} & = & Q_{0}\sum_{j} P_{0j}v_{00j}^{\dag}P_{j}U^{\dag}
          +Q_{0}u^{\dag}\sum_{j}P_{1j}P_{j}U+Q_{1}\sum_{j}P_{1j}\sigma_{xj}
          P_{j}U
          \end{eqnarray}

          $T$ is a sum of three elementary steps: the head in internal
          state $0$ moves to the right and carries out $v_{00}$ on $0$ bits
          only (term 1); if a $1$ bit is encountered, change head state to
          $1$ and move one step to left (term 2); with the head in state
          $1$ shift to the left and flip the encountered bit if and only if
          it is a $1$ (term 3).  

          For most $v_{00}$ in $U(2)$, $T$ is nondeterministic.  $T$ is also
          a partial isometry  (Condition X holds) and it is orthogonality
          preserving. A comparison of $T$ with the example of section
          \ref{Bit-rotation} shows that with the head in state $0$ the
          first term of $T$ is identical with the example. Thus $T$ is
          quantum ballistic on all subspaces in which term $1$ only is
          active and term $2$ is never activated.   These correspond to the
           isometric subspaces of the example in section
          \ref{Bit-rotation}.

          $T$ is also quantum ballistic on subspaces in which term $2$ only
          is active or term $3$ only is active. All paths in the subspaces
          on which term $2$ only is active are of length $1$ (i.e. contain
          two states).  This is an example of the term type with $f(ls)\neq
          l$.  

          However, $T$ does not appear to be quantum ballistic on the
          computation subspace in which all three terms are active.  To see
          this consider the subspace of states $\vert 0\rangle \vert
          l\rangle \vert S_{<l}\rangle \vert S_{\geq l}\rangle$ with $\vert
          S_{<l}\rangle \vert S_{\geq l}\rangle$ given by Eqs. \ref{S<l} and
          \ref{Sgeql}. (The $M$ superscript is suppressed.)  This state 
describes the head at position $l$ and in
          internal state $\vert 0\rangle$ with the spin lattice transformed
          to the left of $l$ and in the $0-1$ spin basis to the right of
          $l$.  Under the action of term $1$, Eq. \ref{Tex1}, the head
          moves ballistically to the right until the $1$ at site $M-1$ is
          encountered.  Then term 2 moves the head back to position $M-1$
          and changes the head state to $\vert 1\rangle$.   Term $3$ now
          becomes active.  However it is active only when it sees a $1$ on
          the transformed component; it annihilates the state when it sees
          a $0$ in the transformed component.  The amplitudes per step for
          following these choices are given by $\langle 1\vert v_{00}\vert
          0\rangle$ and $\langle 0\vert v_{00}\vert 0\rangle$ respectively. 

          This can be stated in another way. Suppose term $1$ is active for
          $n$ steps before a $1$ is encountered where $v_{00}\vert 0\rangle
          =a\vert 0\rangle +b\vert 1\rangle$ with $\vert a\vert ^{2}+\vert
          b\vert ^{2} =1$.  At the end of $n$ steps the lattice spin state
          for the $n$ spins can be expanded as a sum over the $2^{n}$ paths
          (as $0-1$ strings of length $n$) in the computation basis,
          $\sum_{p}c(p)\bigotimes_{l=0}^{n}\vert p(l)\rangle$, with 
          amplitudes $c(p)$ given by $n$ fold products of $a,b$. Iteration
          of term $3$ describes head motion back along each path. As long
          as $1$ are encountered the head moves back to the left along each
          path ballistically until the first $0$ is encountered.  At this
          point the next iteration of term 3 gives $0$, thereby removing
          the path state with its corresponding amplitude from the overall
          system state.  

          This shows that repeated iterations of $T$ generate at some point
          states whose norm starts decreasing below $1$.  This is a result
          of the removal of paths at each step at which term $3$ is active. 
          As the process of iteration of $T$ continues, the amplitude  of
          the remaining state continually decreases. 

It must be emphasized that removal of paths and decrease of overall state 
amplitude refers only to the effect of iterations of $T$ or its adjoint.  It 
does not refer to the actual time evolution of the system.  Since the 
evolution operator $e^{-iHt}$ is unitary no paths are removed and the overall 
state normalization is preserved.  Instead paths 
which are removed by iterations of $T$ correspond to {\em halting} paths of the 
process.  If for some path state $\psi, T\psi =0$, then $\psi$ is the final or 
halting state for the particular path. Dynamically these halting states will be 
seen in future work to act like partially reflecting barriers under the action 
of $e^{-iHt}$.  

          The loss of overall state amplitude under iteration of $T$ or 
          $T^{\dag}$ shows that for this subspace and $T$ given by Eq.
          \ref{Tex1}, the evolution does not proceed quantum ballistically. 
          As defined here quantum ballistic evolution describes either norm 
preserving motion under iteration of $T$ or $T^{\dag}$ or simultaneous
          removal of all paths at the same step.  It does not describe
          motion in which different paths are removed at different
          stages.  

          The above suggests that in order for a nondeterminstic $T$ to be
          quantum ballistic, for a given input string state, all paths in
          the computation basis must be of the same length.  This can be
          achieved by either adding ballast type evolution to each path to
          ensure all paths are infinite, or that all paths are of the same
          finite length.  A simple example of a nondeterministic $T$ with
          all paths of the same finite length is given in Appendix B.  

          It is to be emphasized that the restriction to paths of the same
          length applies only to paths generated as a result of properties
          of $H$. It is not a restriction on the properties of the input
          state.  For example, the input state can be a linear
          superposition of different inputs to the computation, (i.e.
          quantum parallel computation \cite{Deutsch}).  Computation paths
          for each of the component inputs can be of different length
          without affecting the ballistic evolution. This is a consequence
          of the fact that, because the paths on different inputs are
          distinct, the requirement of ballistic evolution applies to each
          input separately.

          \section{Effective Determination of Quantum Ballistic Evolution}
          \label{eff-detn}

          As has been shown in earlier sections, an arbitrary QTM step
          operator defined by Eqs. \ref{Tturing} and \ref{Tls} may not be a
          step operator for a quantum ballistic computation.  In general
          $T$ may be quantum ballistic on some subspace, or not on any
          subspace, or on the whole Hilbert space.
          The question arises of how one determines if $T$ and the
          associated Hamiltonian of Eq. \ref{ham} describe a quantum
          ballistic computation at least on the subspaces of the
          computation. This problem applies to step operators for arbitrary
          process, not just quantum Turing machines. 

          Theorem 5 shows that this question is equivalent to that of the
          existence of an effective decision procedure for determining if a
          step operator is a partial isometry and is stable and
          orthogonality preserving for some basis.  The examples of Section
          \ref{examples}, and many of QTM step operators, show that step
          operators exist which satisfy these requirements. 

          By  "effective procedure" is meant a decision process for
          determining whether or not there exists a basis for which a step
          operator $T$ is orthogonality preserving and stable where the
          number of steps in the decision process is finite. If a numerical
          measure of the complexity of $T$ is available, it is then
          desirable that the number of steps in an effective procedure is
          of the order of a polynomial in the complexity of $T$.  If $T$ is
          a quantum Turing machine step operator, then for the purposes of
          this paper, the complexity of $T$ is of the order of the (finite)
          number of elementary step operators in $T$.

          In particular a decision process is {\em not} effective if it
          requires raising $T$ and $T^{\dag}$ to all positive powers in
          order to make the determination.  The reason is that such a
          process has an infinite number of steps.

          It is clear that for QTMs, one can determine effectively if $T$
          (and its adjoint) is a partial isometry and is orthogonality
          preserving.  This follows from Theorem 1, Eq. \ref{Tls}, and the
          fact that determination if $T^{\dag}T$ and $TT^{\dag}$ commute
          and are equal to their squares requires a number of steps of the
          order of the fourth power of the number of elementary steps in
          $T$.

          The main problem is the effective determination of stability for
          a given step operator. For {\em deterministic} QTM step operators
          such a procedure exists. This follows from the fact that the only
          spin transformations allowed in the terms of the step operator
          $T$ are lattice spin flips or changes of the spin projections of
          the head with respect to a fixed quantization axis.  Simple
          inspection of each of the elementary step terms of the step
          operator $T$ is sufficient to determine if this is the case. 
          Furthermore step operators for deterministic QTMs are usually
          constructed so that the usual computation basis is stable for
          $T$.

          It follows that for deterministic QTM there exists an effective
          decision procedure for deciding if $T$ is orthogonality
          preserving and stable for some basis.

          For nondeterministic QTMs the above proof fails because there 
          does not appear to be a way to determine effectively from
          properties of $T_{ls}$, Eq. \ref{Tls} if $T$ is stable for some
          basis. This can be seen from the arguments of the last section,
          that  show for a given input string state, a search must be
          carried out through on all state paths in the computation basis
          generated by successive iteration of $T$ and $T^{\dag}$ on the
          input state.  If two paths of different length are encountered,
          the computation is not quantum ballistic on the given input.  
          However all iterative powers must be searched before one can
          conclude a computation is quantum ballistic 

          It is thus concluded that an effective decision procedure exists
          for determining for the step operator for an arbitrary
          deterministic quantum Turing machine if a simple Hamiltonian
          description of quantum ballistic evolution exists for the
          computation.  It is an open question if such a procedure exists
          for the step operator of a nondeterministic quantum Turing
          machine. 

          Note that, given a step operator for a nondeterministic QTM, one
          can use some prescription such as that provided by Bennett
          \cite{Bennett} to add history and copying degrees of freedom. 
          The problem is to determine for each input to the computation, if
          there exists a basis for which the expanded step operator $T$ is
          stable on the computation subspace (the subspace spanned by the
          states obtained by all iterations of $T$ and $T^{\dag}$ on the
          input state. As noted before an effective procedure exists for
          determining if $T$ is orthogonality preserving. 

          It is of interest to consider the case in which a process step
          operator $T$ is a power partial isometry and has no unitary
          components, (except possibly for bilateral shifts and cyclic
          finite orbits). In this case the main theorem, Theorem 5, holds
          with orthogonality preservation and stability replaced by
          complete orthogonality preservation.  In this case the decision
          process involves determining if a step operator $T$ is a partial
          isometry and is completely orthogonality preserving (Theorem 4). 

          It is easy to determine effectively if $T$ is a partial isometry:
          one needs only to determine if $I_{1}$ and $F_{1}$ are projection
          operators (i.e. if $(I_{1})^{2}=I_{1}$ and $(F_{1})^{2})=F_{1}$). 
          The problem is to determine effectively if $T$ is completely
          orthogonality preserving.

          The nonexistence of an effective decision procedure for complete
          orthogonality preservation follows from the Halmos-Wallen
          counterexample. It shows that one cannot conclude complete
          orthogonality preservation from orthogonality preservation and
          that all positive powers of $T$ and its adjoint would have to be
          inspected. In particular it shows that operators exist for which
          all powers up to some arbitrarily large $n$ are orthogonality
          preserving, but the $n+1st$ is not.  

          \section{Discussion}
          \label{discussion}

          Several aspects of the material presented so far in this paper
          should be stressed.  First, it is important to emphasize the
          distinction between the reversibility and ballistic properties of
          a step operator $T$ and those of the associated Feynman
          Hamiltonian $H=K(2-T-T^{\dag})$.  In general $T$, including those
          defined for QTMs by Eqs. \ref{Tturing} and \ref{Tls} do not even
          describe reversible processes. Or they may describe reversible
          processes which do not evolve ballistically. An example of the
          latter would be any unitary $T$ which is not stable on any basis
          (Theorem 5). 

          An example of the former is the erasure operator $T=\sum_{j=-
          \infty}^{\infty} (P_{0,j}+\sigma_{x,j}P_{1,j})UP_{j}$. which
          describes resetting of all $1's$ in a string to $0's$.  This
          process is not reversible as iteration of $T$ describes paths
          which join.  As Landauer \cite{Landauer} has emphasized,
          information is destroyed.  In this case the Hamiltonian of Eq.
          \ref{ham} will describe evolution of another process which is
          reversible and not that associated with iteration of $T$. 

          The definition of quantum Turing machines used here (Section
          \ref{qctm}) differs from that proposed by Deutsch \cite{Deutsch}
          and Bernstein and Vazirani \cite{BV} and which is often quoted in
          the literature.  These authors restrict $T$ to be unitary and
          local in the computation basis and to apply to a finite time. 
          That is, $T=e^{-iHt}$ for some finite time interval $t$.

          As was noted earlier in this paper, it is impossible to satisfy
          these requirements with a Hamiltonian which is local and is
          simple  in that it has the complexity of $T$ and not of all
          powers of $T$ and $T^{\dag}$.  This suggests that one combine the
          two approaches by restricting $T$ to be unitary and to refer to
          infinitesimal time steps only.  In this way the Feynman
          Hamiltonian \cite{Feynman}, Eq, \ref{ham}, constructed from
          unitary $T$ is local and is simple.

          There are some problems with this approach.  The main problem is that 
it is unnecessarily restrictive to require $T$ to be unitary.  Step operators 
constructed as sums of local step or program elements used in algorithms are 
not likely to be unitary.  The definition given here in which $T$ is not even 
required to be normal is more general and it corresponds more
          closely to what one actually does in construction of algorithms
          as sums of local step or program elements.  As an example
          consider Simon's algorithm \cite{Simon} for a quantum
          computation.  This consists of two "Fourier transforms" separated
          by the computation of a function $f:\{0,1\}^{n}\Rightarrow
          \{0,1\}^{m}$ with $m\geq n$ to generate the state $\sum
          _{p}c(p)\vert p\rangle \bigotimes \vert f(p)\rangle$.\footnote{
          Here oracle presentation of $f$ is replaced by computation of $f$
          to obtain a physically meaningful procedure.}  The sum is over
          all $2^{n}$ $0-1$ strings $p$ of length $n$.  It is clear that
          any step operator $T$, which corresponds to a sum of local steps
          for computation of the function and generation of $\vert p\rangle
          \bigotimes \vert f(p)\rangle$ from $\vert p\rangle$ is not likely
          to be unitary.

          Another way around this problem might be to consider any step
          operator $T$ for a QTM as defined here such that $T$ is a partial
          isometry and is orthogonality preserving on some basis.  Then
          dilate $T$ to a unitary operator on the whole Hilbert space by
          suitable extension of the definition of $T$ to the null subspace. 
          One problem with this is that there is no way to effectively
          define either the null subspace or the subspace on which $T$ is 
unitary. 

This problem was already mentioned for earlier work \cite{Benioff1,Benioff2}.
In particular, the null (or unitary) subspace 
          consists of all states in some basis which are not (or are) reached at
          some stage of the computation on some input.  If $T$
          is the step operator for a (deterministic) universal Turing
          machine, an effective definition of the null or unitary subspaces 
would require solution of the halting problem which is impossible.

          Another problem with this approach is that even if the dilation
          is unitary, powers of the dilation would not correspond to powers
          of the original $T$.  This was examined elsewhere \cite{Benioff3}
          where minimal unitary power dilations $V^{T}$ of step operators
          were constructed.  It was seen that the construction added an
          extra degree of freedom and that history was generated
          automatically when needed. 

          However examples of deterministic Turing machine step operators
          showed that $T$ needed to be expanded by addition of history and
          copy degrees of freedom  prior to unitary power dilation.  This
          was needed to avoid most of the state amplitude going into
          history components.  This suggests that, at least for
          deterministic QTMs, unitary power dilation with the addition of
          another degree of freedom is not needed.  It is not known if
          unitary power dilation of step operators for nondeterministic
          QTMs has any advantages.

          Another problem with expansion of a process step operator into a
          unitary operator by addition of degrees of freedom is that all
          degrees of freedom need to be kept isolated from the environment
          so that coherence between phases of the states of all degrees of
          freedom are maintained. This is especially important for
          constructing quantum mechanical computers as their operation (for
          example, Shor's algorithm \cite{Shor}) depends on maintaining
          phase relations among the different components.

          This suggests that it is important to minimize the number of
          degrees of freedom to be added.  In this way effects of the
          environment, such as decoherence, etc.
          \cite{Unruh,C-L-S-Z,Cald-Shor,PGCZ} make it useful to minimize
          the number of additional degrees of freedom that need to be
          protected. 

          The results of this paper suggest that to ensure quantum
          ballistic evolution, it is sufficient to add just enough
          additional degrees of freedom so that a step operator $T$ for the
          expanded process is a partial isometry, preserves orthogonality,
          and is stable for some basis.  In particular, it is not necessary
          to add even more degrees of freedom to ensure that the expanded
          operator is unitary.  If $T$ is stable and orthogonality
          preserving in some basis, then (Theorem 5) for such processes
          there always exists a simple time independent Hamiltonian (for
          example that of Eq. \ref{ham}) which correctly models quantum
          ballistic evolution  a simple time independent Hamiltonian (for 
example that of Eq. \ref{ham}) which correctly models quantum ballistic 
evolution. It is, however, an   open question how one can effectively 
determine the minimal number of degrees of freedom needed to guarantee 
reversible or quantum ballistic evolution.
          was limited here to quantum Turing machines.  It also applies to
          other models of quantum computation such as quantum circuits. 
          Specfically, any quantum circuit which can be modelled by a step
          operator and for which quantum ballistic evolution is a
          satisfactory description of the computation, is included.

          \section{Future Work}
          Much of the concern of this paper was with necessary and
          sufficient conditions for a step operator $T$ to generate quantum
          ballistic evolution for a process. It was seen that if $T$ was a
          partial isometry, orthogonality preservation and stability gave
          for the Feynman Hamiltonian, Eq. \ref{ham} a canonical form for
          both the eigenvalues and eigenfunctions and the description of
          quantum ballistic evolution.  

          As was seen there are many processes which fit these
          requirements.  However there are also many processes in physics
          with associated step operators which are reversible but do not
          evolve quantum ballistically.  The work of this paper needs to be
          generalized to accomodate these.  Also the consequences of
          orthogonality preservation and stability for step operators which
          are not partial isometries needs to be investigated.

          Also it was shown that it is an open question if there exists an
          effective decision procedure to determine if a step operator for
          a process such as a nondeterministic QTM describes quantum
          ballistic evolution.  This open question needs to be closed,
          either by giving an effective decision procedure or by proof that
          the question is effectively undecidable.

          \section*{Acknowledgements}
          The author would like to thank Dr. Fritz Coester for providing
          valuable review comments for this work.  This work is supported
          by the U.S. Department of Energy, Nuclear 
          Physics Division, under contract W-31-109-ENG-38.

          \newpage
          \begin{appendix}
          \begin{center}
          {\bf APPENDIX A}
          \end{center}

          {\bf Theorem 1} {\em An operator $T$ and its adjoint are
          orthogonality preserving if and only if $T^{\dag}T$ and
          $TT^{\dag}$ commute.} \\
          \\
          Proof:
          Let $\{\vert n\rangle :n=0,1,\cdots \}$ denote a common basis set
          for which both $T$ and $T^{\dag}$ are weakly orthogonality
          preserving.  Then
          \begin{eqnarray*}
          & \langle n\vert T^{\dag}TTT^{\dag} - TT^{\dag}T^{\dag}T\vert 
          m\rangle \\
          & = \sum_{j}[\langle n\vert T^{\dag}T\vert j\rangle \langle
          j\vert TT^{\dag}\vert m\rangle - \langle n\vert TT^{\dag}\vert
          j\rangle \langle j\vert T^{\dag}T\vert  m\rangle.
          \end{eqnarray*}

          By the definition of weak orthogonality preservation the
          component matrix elements are different from $0$ only if $j=m$
          and $j=n$.  This is impossible if $m\neq n$.

          Conversely suppose $T^{\dag}T$ commutes with $TT^{\dag}$.  Since
          both these operators are self adjoint, by the spectral theorem,
          there exist two spectral measures. ${\cal E}$ and ${\cal F}$ such
          that 
          \begin{eqnarray*}
          T^{\dag}T & = & \int \lambda d{\cal E}_{\lambda} \\
          TT^{\dag} & = & \int \lambda d{\cal F}_{\lambda}.
          \end{eqnarray*}
          Since $T^{\dag}T$ and $TT^{\dag}$ commute, there exists another
          spectral measure ${\cal G}$ which is a common refinement or
          product of ${\cal E}$ and ${\cal F}$.  Let $\{\vert r\rangle$:
          $r$ is in the spectrum of $T^{\dag}T$ or $TT^{\dag} \}$  be a
          basis set of continuous or discrete eigenfunctions defined by
          ${\cal G}$.  Here physicists license is being used to  speak of
          continuous eigenfunctions.  In case of degeneracy, extra basis
          labels are implicitly assumed.

          By construction it is clear that if $r\; \epsilon$ spectrum
          $T^{\dag}T$, then ${\cal E}_{s} \vert r\rangle = \vert r\rangle
          \; [0]$ if $s>r \; [s\leq r]$.  If $r$ is not in the spectrum of
          $T^{\dag}T$, then ${\cal E}_{s} \vert r\rangle =0$ for all $s$. 
          Similar relations hold for ${\cal F}$ and $TT^{\dag}$.

          It follows that if $r\neq s$ then $\langle s\vert T^{\dag}T\vert
          r\rangle = \langle s\vert TT^{\dag}\vert r\rangle =0$, which
          proves the theorem.\\
          \\
          {\bf Theorem 3} {\em A partial isometry $T$ is orthogonality
          preserving and stable in some basis $B$ if and only if $T$ is
          distinct path generating in $B$.} \\
          \\
          Proof:$\Longrightarrow$: If $T$ and $T^{\dag}$ are orthogonality 
preserving and stable in a
          basis $B$ then iteration of $T$ or its adjoint generates paths in
          $B$.  This follows from the definition of stability and bases as
          $T\vert p_{i}\rangle$ is in $B$ if $\vert p_{i}\rangle$ is in $B$
          and $T\vert p_{i}\rangle \neq 0$.

          To show that $T$ and $T^{\dag}$ are distinct path generating
          suppose that two states $\vert p_{i}\rangle$ and $\vert
          p_{j}\rangle$ are in different paths as generated by iterations
          of $T$ or its adjoint and that there exist smallest values $m,n$
          such that $\langle T^{n}p_{j}\vert T^{m}p_{i}\rangle \neq 0$.
          That is starting from $\vert p_{i}\rangle$ and $\vert
          p_{j}\rangle$, the paths first intersect after $m$ and $n$
          iterations of $T$ respectively. By assumption  $\langle T^{n-
          1}p_{j}\vert T^{m-1}p_{i}\rangle =0$. But orthogonality
          preservation and stability implies that $\langle T^{n}p_{j}\vert
          T^{m}p_{i}\rangle =\langle T(T^{n-1})p_{j}\vert T(T^{m-
          1})p_{i}\rangle =0$ which is a contradiction.  Thus $T$ is
          distinct path generating.  Repetition of the above for $T^{\dag}$
          proves the implication.

          $\Longleftarrow$:  Assume $T$ and $T^{\dag}$ are distinct path
          generating in some basis $B$.  From the definition of distinct
          path generation, stability in $B$ follows immediately. 
          Orthogonality preservation also follows:  to see this assume
          first that $\vert p_{i}\rangle , \vert p_{j}\rangle$ are distinct
          states in the same path for $T$. By the definition of a path,
          either $T\vert p_{i}\rangle =0$, $T\vert p_{j}\rangle =0$ or both
          are different from $0$ and are different states.  In all these
          cases $\langle Tp_{i}\vert Tp_{j}\rangle =0$.  

          If $\vert p_{i}\rangle$ and $\vert p_{j}\rangle$ are in different
          paths, then by assumption $T\vert p_{i}\rangle$ and $T\vert
          p_{j}\rangle$ are either $0$ or are distinct.  Thus  $\langle
          Tp_{i}\vert Tp_{j}\rangle =0$ which shows that $T$ is
          orthogonality preserving.  Repetition the above for $T^{\dag}$
          completes proof of the theorem.\\
          \\
          {\bf Theorem 4} {\em A partial isometry $T$ and its adjoint are
          completely orthogonality preserving if and only if $T$ is a power
          partial isometry.} \\
          \\
          Proof: $\Longrightarrow$ :
          For each $n$ let the operators $I_{n}$ and $F_{n}$ be defined by
          Eqs. \ref{In} and \ref{Fn}. Since complete orthogonality
          preservation implies orthogonality preservation, Theorem 1
          implies that for each $n$, $[I_{n},F_{n}]=0$.   

          Claim: for all $n,m$ with $n$ different from $m$,
          $[I_{n},F_{m}]=0$.  To see this let $\{\vert p_{j}\rangle\}$ be
          the common basis which preserves orthogonality for all powers of
          $T$ and its adjoint.  By hypothesis and the definition of
          complete orthogonality preservation, such a basis exists. One
          also has
          \begin{eqnarray*}
          & \langle p_{i}\vert I_{n}F_{m} - F_{m}I_{n}\vert p_{l}\rangle \\
          & = \sum_{j}[\langle p_{i}\vert I_{n}\vert p_{j}\rangle \langle
          p_{j}\vert F_{m}\vert p_{l}\rangle - \langle p_{i}\vert
          F_{m}\vert p_{j}\rangle \langle p_{j}\vert I_{n}\vert 
          p_{l}\rangle.
          \end{eqnarray*}
          Since $\langle p_{i}\vert I_{n}\vert p_{j}\rangle$ and $\langle
          p_{i}\vert F_{m}\vert p_{j}\rangle =0$ if $i\neq j$, the above
          expression is $0$ since for $i\neq l$, $j$ cannot be equal to
          both $i$ and $l$.  For $i=l$ the expression is clearly equal to
          $0$. Thus $[I_{n},F_{m}]=0$.

          The final step is by induction.  One already has that $T$ and
          $T^{2}$ are partial isometries.  Assume that $T^{n}$ and $T$ are
          partial isometries. Then $T^{n+1}$ is a partial isometry.  This
          follows from the above proof that $[I_{1},F_{n}]=0$, and the H-W
          lemma.

          $\Longleftarrow$:
          Since $T$ is a power partial isometry, it can be decomposed
          [Halmos-Wallen] \cite{Halmos-wallen} into a unitary operator on
          the subspace ${\cal H}_{1}=F_{\infty}I_{\infty}\cal H$, an
          isometry on ${\cal H}_{2}=I_{\infty}-I_{\infty}F_{\infty}\cal H$,
          a coisometry on ${\cal H}_{3}=F_{\infty}-I_{\infty}F_{\infty}\cal
          H$, and for each $n$ a truncated shift of index $n$ on ${\cal
          H}_{4,n}=P_{n}\cal H$.

          To prove the implication it is necessary to show the existence of
          a basis for each of the reducing components for which $T$ is
          completely orthogonality preserving.  For the unitary part any
          selected basis will do because $T^{\dag}T=TT^{\dag}=1$ on ${\cal
          H}_{1}$.  That is, for any basis on this subspace  $\{\vert
          n\rangle\},\: n\neq m\rightarrow \langle T^{l}n\vert
          T^{l}m\rangle = \langle (T^{\dag})^{l}n\vert
          (T^{\dag})^{l}m\rangle =0$  for $l=0,1,\cdots$.

          For the isometric component use is made of the fact that any
          isometry is a direct sum of a unitary part and copies of
          unilateral shifts \cite{Halmos2}.  The unitary part is included
          above.  For the unilateral shifts select for the basis the set
          $\{\vert n,k\rangle\}$.  Here $k$ is the index representing a
          term in the direct sum and $T\vert n,k\rangle =\vert
          n+1,k\rangle$ for $n=0,1,\cdots$.  It is clear that complete
          orthogonality preservation occurs for this basis since, if $n\neq
          m$ then for all $j$, $\langle T^{j}n,k\vert T^{j}m,k\rangle
          =\langle n+j,k\vert m+j,k\rangle =0$ and $\langle
          (T^{\dag})^{j}n,k\vert (T^{\dag})^{j}m,k\rangle =\langle n-
          j,k\vert m-j,k\rangle =0$.  The last equality for the adjoint of
          $T$ is trivially true for $n-j<0,m-j<0$.

          For the coisometric component the above argument can be repeated
          by exchanging $T$ with its adjoint and letting $n$ range over the
          nonpositive integers.

          For the truncated shifts of index $n$, the argument given for
          isometries can be repeated.  That is, the operator $T_{n}$ which
          is the restriction of $T$ to the reducing subspace ${\cal
          H}_{4,n}=P_{n}\cal H$ can be written as a direct sum
          $\oplus_{k}T_{k,n}$ where $T_{k,n}$ is a truncated shift on the
          kth component of ${\cal H}_{4,n}$.  

          Halmos and Wallen \cite{Halmos-wallen} (see also
          \cite{Hoover-lambert}) have shown that the projection operator
          $P_{n}$ can be defined by the orthogonal sum
          \begin{equation}
          P_{n}=\sum_{l=1}^{n}\Delta I_{n-l}\Delta F_{l} 
          \end{equation}
          where $\Delta I_{n-l}=I_{n-l}-I_{n-l+1}$ and $\Delta F_{l}=F_{l-
          1}-F_{l}$.  The $I$ and $F$ projection operators are defined by
          Eqs. \ref{In} and \ref{Fn}. 

          Let $\{\vert j,k\rangle\}$ represent a basis on ${\cal H}_{4,n}$
          such that for each $l$,
          \[ T^{l}\vert j,k\rangle = \left\{\begin{array}{ll}
          \vert j+l,k\rangle & \mbox{if $j+l\leq n$} \\
          0 & \mbox{if $j+l>n$}. 
          \end{array} \right. \]
          One also has
          \[ (T^{\dag})^{l}\vert j,k\rangle = \left\{\begin{array}{ll}
          \vert j-l,k\rangle & \mbox{if $j-l\geq 0$} \\
          0 & \mbox{if $j-l<0$}. 
          \end{array}
          \right. \]  It is clear from the above that $T_{n}$ and its
          adjoint are completely orthogonality preserving on the defined
          basis.  Since all cases of the decomposition of $T$ are covered,
          the proof of the theorem is complete.  

          \newpage
          \begin{center}
          {\bf APPENDIX B}
          \end{center}
          The goal here is to exhibit an example of a nondeterministic QTM
          step operator $T$ which is quantum ballistic. This is done by
          ensuring that for each input all computation paths are the same
          length.  For the purposes of illustration the example will be
          made simple, with only one nondeterministic step. 

          Define a step operator $T$ by
          \begin{eqnarray}
          T & = & Q_{0}\sum_{j}P_{0j}UP_{j} (1) + uQ_{0}\sum_{j}v_{j}
          P_{1j}U^{\dag}P_{j} (2) + uQ_{1}\sum_{j}P_{0j}UP_{j} (3) \nonumber  \\
          &  & \mbox{}+ uQ_{2}\sum_{j}P_{0j}UP_{j} (4)+ 
           u^{2}Q_{2}\sum_{j}P_{1j}UP_{j}(5)  \label{Tex2}
          \end{eqnarray}
          The component operators are defined here as before.  Recall that
          $v_{j}$ is any unitary operator in $U(2)$ and is the same for
          each $j$.  The numbers in parentheses following each term are
          included for easy reference and are not part of the equation.

          The first term moves the head in state $0$ to the right along a
          string of lattice $0s$.  Term $2$ carries out the only
          nondeterministic step by applying a $v$ transformation to the
          first $1$ encountered, changing the head state to $1$ and moving
          one step back.  Term $3$ moves the head back to the transformed
          bit, changing the head state to $2$.  The next two terms generate
          a path split in the computation basis by moving the head one step
          to the right and changing the head state to a $3$ or a $4$ if $0$
          or $1$ is encountered respectively.  The  computation then stops
          after producing two paths, each of length $1$, after the split.  
          It is a straight forward but tedious exercise to show that $T$ is
          a power partial isometry.  Note that it is sufficient to examine
          all powers of $T$ and $T^{\dag}$ up to the fourth since all
          higher powers have the same structure as the
          fourth\footnote{$T^{4} =1111+2111+3211+4321+5321$. Here the
          single digits denote the term numbers in Eq. \ref{Tex2} and the
          order of the digits gives the order in which each of the $T$
          terms appears.  Higher powers of $T$ just add more $"1"$ terms to
          the right of each of the $5$ terms of $T^{4}$.}.  Thus the
          Halmos-Wallen decomposition applies and $T$ can be decomposed
          into unitary, isometric, coisometric, and finite truncated shift
          components.

          It remains to show that $T$ is stable on some basis.  This will
          be done by explicit construction of the basis in the subspaces
          associated with each of the components. The unitary component is
          limited to the subspace spanned by the basis $\vert 0,j,s\rangle$
          for all $j$ where $\vert s\rangle = \bigotimes _{k=-
          \infty}^{\infty} \vert 0\rangle_{k}$ is the constant $0$ sequence
          on the lattice.  On this subspace, the first term of $T$ in Eq.
          \ref{Tex2} is the only active term.

          For most of the remaining components it suffices to consider a
          finite section of the lattice consisting of $n\: 0s$ terminated
          on both ends by $1s$.  That is $\vert s\rangle =\vert
          1\rangle_{M} \bigotimes_{k=M+1}^{k=L} \vert 0\rangle_{k}
          \bigotimes \vert 1\rangle_{L+1} = \vert s'\rangle \bigotimes
          \vert 1\rangle_{L+1}$ where $L-M=n$.  For all head positions $k$
          between $M+1$ and $L$ in the state $\vert 0,k,s\rangle$, only
          term $1$ is active in the iteration of $T$ or $T^{\dag}$ moving
          the head to either end of the lattice segment.  Terms $2$ and $3$
          acting in successtion convert the state $\vert 0, L+1,s\rangle$
          into $\vert 1,L,s'\rangle v\vert 1\rangle_{L+1}$ into $\vert
          2,L+1,s'\rangle v\vert 1\rangle_{L+1}$.

          Both terms $4$ and $5$ are active in the next iteration of $T$. 
          The state generated is $(a\vert 3\rangle \vert 0\rangle_{L+1}
          +b\vert 4\rangle \vert 1\rangle_{L+1})\vert L+2,s'\rangle$ where
          $a=\langle 0\vert v\vert 1\rangle, b=\langle 1\vert v\vert
          1\rangle$. The state shows the path split in the computation
          basis with the head state $3$ correlated with a $0$ bit at site
          $L+1$ and the head state $4$ correlated with a $1$ bit at the
          site.

          The next iteration of $T$ annihilates both components of the
          above state giving the truncation at one end.  Thus the two paths
          are of the same length. The states listed above along with
          similar ones obtained by iteration of $T^{\dag}$ form a stable
          basis for a truncated shift of length $n+7$.  Note that there are
          an infinite number of copies of this shift since there are an
          infinite number of basis states spanning the lattice region
          outside the interval $[M\geq n\geq L+1]$. Also the $[M-L]$
          interval can be shifted to any position on the lattice. 

          The above description, applied to each value of $n$, gives a
          description of the stable basis for all truncated shifts of
          length $8$ or more.  Setting $L =\infty$ or $M=-\infty$ gives the
          stable basis for the respective isometric and coisometric
          components.  Stable basis states for the truncated shift
          components of length $<8$ can also be easily described.

          The above explicit description of a stable basis and the fact
          that $T$ is a power partial isometry show (Theorem 5) that the
          Hamiltonian of Eq. \ref{ham} describes quantum ballistic
          evolution for $T$ even though it is nondeterministic.  In this
          case the fact that $T$ is a power partial isometry is sufficient
          proof since the unitary part is a bilateral shift.

          For the step operator as defined each path has length $1$ after
          the split.  It is easy to extend the definition of $T$ so that
          the paths have length $n$ for any $n$.  The definition can also
          be extended so that $T$ is quantum ballistic on some computation
          subspaces and not on others. This is the case if the terms
          $Q_{3}\sum_{j}P_{1j}UP_{j}$ and $Q_{4}\sum_{j}P_{0j}UP_{j}$ are
          added to $T$ defined by Eq. \ref{Tex2}.
          \end{appendix}

          \newpage
                    \begin{center}
                    FIGURE CAPTIONS
                    \end{center}

          Figure 1.  Schematic Representation of Quantum Ballistic Paths. 
          The points (as solid circles) represent different basis states in
          a given basis.  The coordinate distance and relative location of
          the points in the x-y plane has no meaning and is given for
          illustrative purposes only.  Two infinite paths are shown with
          dashed lines.  Path A shows no terminus and path B terminates at
          state T (no relation to the step operator T).  The coefficients for 
each of two wave packets $\psi_{1}(t),\psi_{2}(t)$ are shown as short vectors 
at each point on the paths. The coefficients $c_{n}(t)$ (Eq. \ref{packet}) are
          shown in the figure as $c_{n}(t) = r(t)e^{i\theta (t)}$ where polar 
coordinates are used.  The n-dependence of $r(t)$ and $\theta(t)$ are shown 
explicitly.
\\
            
          Figure 2. The Lattice State for Bound State Motion in the Presence of 
$0s$, Section VI-A.  The figure shows $1s$ at $N$ and $N+W+1$ and $0s$
          elsewhere. The solid vertical lines denote the positions at which the 
bound states are $0$  (Eq. \ref{eq:b-eigens0}).  As such they correspond to 
completely reflecting barriers. 
          
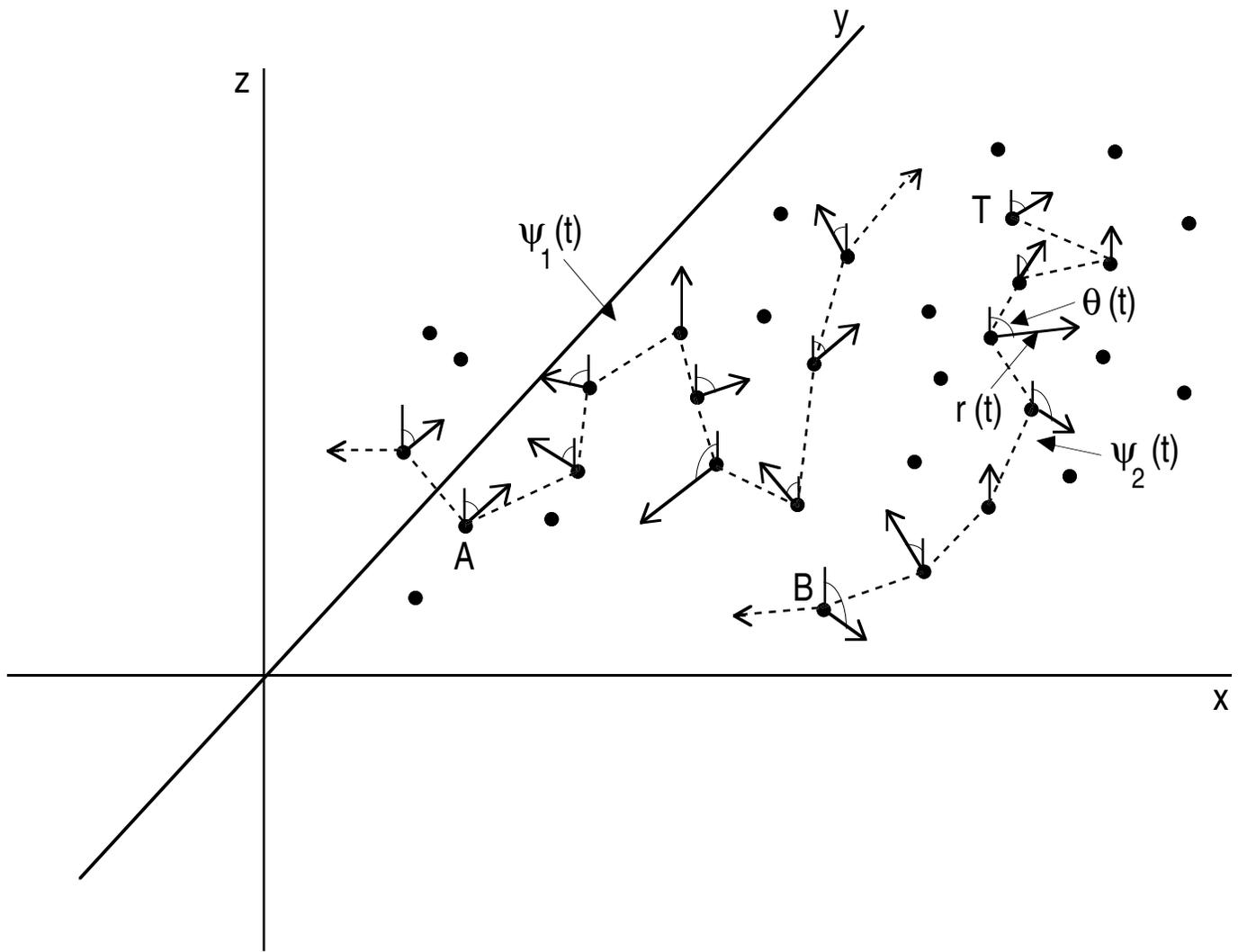
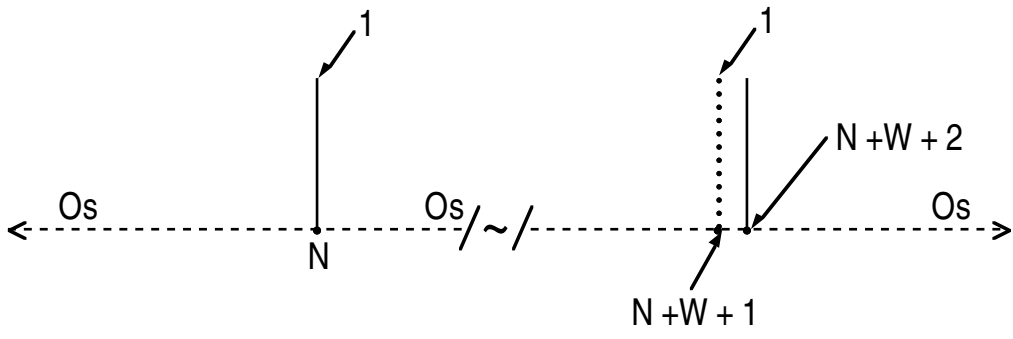
\end{document}